\documentclass[journal]{IEEEtran}

\usepackage{graphicx}
\usepackage{color}
\usepackage{amsfonts}
\usepackage{float}
\usepackage{amsmath}
\usepackage{amsthm}
\usepackage{epstopdf}
\usepackage{latexsym}
\usepackage{mathrsfs}
\usepackage{setspace}
\usepackage{changepage}
\usepackage{relsize}
\usepackage{scalefnt}
\usepackage{array}
\usepackage{cite}
 \usepackage{relsize}
\newcolumntype{L}[1]{>{\raggedright\let\newline\\\arraybackslash\hspace{0pt}}m{#1}}
\newcolumntype{C}[1]{>{\centering\let\newline\\\arraybackslash\hspace{0pt}}m{#1}}
\newcolumntype{R}[1]{>{\raggedleft\let\newline\\\arraybackslash\hspace{0pt}}m{#1}}

\let\EndItemize\enditemize
\def\enditemize{\EndItemize\bigskip}
\newcommand*{\LargerCdot}{\raisebox{-0.25ex}{\scalebox{1.5}{$\cdot$}}}

\newcommand{\tabincell}[2]{\begin{tabular}{@{}#1@{}}#2\end{tabular}}

\begin{document}

\title{Estimating the Propagation of Interdependent Cascading Outages With Multi-Type \\ Branching Processes}

\author{Junjian~Qi,~\IEEEmembership{Member,~IEEE,}
        Wenyun Ju,~\IEEEmembership{Student Member,~IEEE,}
        and Kai Sun,~\IEEEmembership{Senior Member,~IEEE}
        \thanks{This work was supported in part by U.S. Department of Energy, Office of Electricity Delivery and Energy Reliability and the CURENT Engineering Research Center. Paper no. TPWRS-01531-2015.
        
        J.~Qi is with the Energy Systems Division, Argonne National Laboratory, Argonne, IL 60439 USA (e-mail: jqi@anl.gov).
        
        W. Ju and K. Sun are with the Department of Electrical Engineering and Computer Science, University of Tennessee, Knoxville, TN 37996 USA (e-mail: wju1@vols.utk.edu; kaisun@utk.edu).
}
}

\markboth{Preprint of DOI 10.1109/TPWRS.2016.2577633, IEEE Transactions on Power Systems.}{stuff}
\maketitle

\begin{abstract}
In this paper, the multi-type branching process is applied to describe the statistics and interdependencies of line outages, the load shed, and isolated buses. 
The offspring mean matrix of the multi-type branching process is estimated by the Expectation Maximization (EM) algorithm and can quantify the extent of outage propagation. 
The joint distribution of two types of outages is estimated by the multi-type branching process via the Lagrange-Good inversion. 
The proposed model is tested with data generated by the AC OPA cascading simulations on the IEEE 118-bus system. 
The largest eigenvalues of the offspring mean matrix indicate that the system is closer to criticality when considering the interdependence of different types of outages. 
Compared with empirically estimating the joint distribution of the total outages, good estimate is obtained by using the multi-type branching process with a much smaller number of cascades, thus greatly improving the efficiency. It is shown that the multi-type branching process can effectively predict the distribution of the load shed and isolated buses and their conditional largest possible total outages even when there are no data of them. 
\end{abstract}


\section{Introduction}

\IEEEPARstart{L}{arge} and rare cascading blackouts are complicated sequences of dependent outages that successively weaken a power system. 
They have substantial risk and pose great challenges in simulation, analysis, and mitigation \cite{usblackout,nerc,TF}. 
General cascading failures have been studied in abstract network models, such as the  Motter-Lai model \cite{Motter,zhao} and the sandpile model \cite{sandpile}.
Specifically for electric power systems, simulations of cascading outages from various models, such as OPA\footnote{OPA stands for Oak Ridge National Laboratory, Power Systems Engineering
Research Center at the University of Wisconsin, University of Alaska to indicate the institutions collaborating to devise the simulation.} model \cite{OPA1,OPA2,OPA3,OPA4}, 
AC OPA \cite{ACOPA,ACOPA1}, OPA with slow process \cite{slow}, Manchester model \cite{kirschen}, hidden failure model \cite{thorp}, \cite{chen}, and dynamic model \cite{dynamic}, can produce massive amounts of data regarding line outages, generator tripping, and load shedding. However, simulations cannot produce statistical insight or metrics with actionable information 
without a carefully designed information extraction method.

Existing such methods include the interaction network and interaction model \cite{interaction,interaction1}, the influence graph \cite{linegraph}, and the branching processes \cite{Ian10,Kim12,bp13,Ren08,Ian12}. Among these methods, branching processes have descriptive parameters that characterize the system resilience to cascading. It is much more time efficient by first estimating the parameters of a branching process from a shorter simulation run and then predicting the distribution of total outages using the branching process than empirically estimating the distribution. Branching processes can efficiently predict the distribution of line outages and the load shed of simulations from OPA and its variants on IEEE 118- and 300-bus systems, and Northeastern Power Grid of China \cite{Ian10,Kim12,bp13}, the distribution of the load shed for the TRELSS simulation on an industrial system of about 6250 buses \cite{Kim12}, and the distribution of line outages in real data \cite{Ren08,Ian12}.

Till now the branching process has only been used to describe the propagation of one type of outages. 
In real cascading blackouts, however, several outages such as line outages, load shedding, and isolated buses can exist simultaneously. 
More importantly, these outages are usually interdependent and thus their propagation can be better understood only when they can be described jointly. 
Also, if we want to evaluate the time that is needed to restore the system after a cascading outage event, we need to know how many
buses and lines are still in service, as well as the amount of the load shed. 
But we may not have all these data and thus need to predict some of them by only using the available data. 

In this paper, line outages, the load shed, and isolated buses, are described by Galton-Watson multi-type branching processes \cite{KB:04,TE:89}. 
The parameters of branching processes are estimated by the Expectation Maximization (EM) algorithm \cite{em}. 
The joint distributions of total outages are efficiently estimated by multi-type branching processes via the Lagrange-Good inversion \cite{good}. 
We also show that the multi-type branching process can effectively predict the distribution of the load shed and isolated buses and their conditional largest possible total outages when there are no data for them. 

Note that the multi-type branching process discussed in this paper can not only quantify the interdependencies between different types of outages in power systems, 
but can also be used to study the interactions between different infrastructure systems, such as between electric power systems and communication networks \cite{inter,commu}, 
natural gas networks \cite{gas1,gas}, water systems \cite{water}, and transportation networks \cite{transport,transport1}.

The rest of this paper is organized as follows. Section \ref{G-W BP} briefly introduces the multi-type Galton-Watson branching processes.
Section \ref{estimating} explains the estimating of branching process parameters.
Section \ref{total pdf} discusses the estimation of the joint distribution of total outages by multi-type branching processes.
Section \ref{num needed} determines how many cascades should be simulated and how many should be utilized to estimate the joint distribution by branching processes. 
Section \ref{simulation} tests the proposed method with simulated cascades of line outages, the load shed, and isolated buses by AC OPA on the IEEE 118-bus system. Finally the conclusion is drawn in Section \ref{conclusion}.

\section{Galton-Watson Branching Processes}
\label{G-W BP}

Here, we will briefly introduce the Galton-Watson branching process, especially the multi-type Galton-Watson branching process. 
For more details, the reader is referred to \cite{Ian10,Kim12,bp13,Ren08,Ian12,KB:04,TE:89}. 

As a high-level probabilistic model, the branching process can statistically describe how the number of outages propagates in a cascading blackout and the statistics of the total number of outages, 
which is different from the OPA model and its variants \cite{OPA1,OPA2,OPA3,OPA4,ACOPA,ACOPA1,slow} that retain information about the network topology, power flow, and the operator's response, 
or the interaction model \cite{interaction,interaction1} that aims at quantifying the interactions between component failures.
The simplicity of the branching process allows a high-level understanding of the cascading process without getting entangled in the complicated mechanisms of cascading.

For one-type branching process, the initial outages propagate randomly to produce subsequent outages in generations. 
Each outage (a ``parent'' outage) independently produces a random nonnegative integer number of outages (``children'' outages) in the next generation. 
The children outages then become parents to produce another generation until the number of outages in a generation becomes zero.

The distribution of the number of children from one parent is called the \textit{offspring distribution}. 
The mean of this distribution is the parameter $\lambda$, which is the average number of children outages for each parent outage and 
can quantify the tendency for the cascade to propagate in the sense that larger $\lambda$ corresponds to faster propagation. 
For cascading blackout $\lambda<1$ and the outages will always eventually die out.

The multi-type branching process is a generalization of the one-type branching process. 
Each type $i$ outage in one generation (a type $i$ ``parent'' outage) independently produces a random nonnegative integer number of outages of the same type (type $i$ ``children" outages) 
and any other type (type $t$ ``children" outages where $t \neq i$). 
All generated outages in different types comprise the next generation. The process ends when the number of outages in all types becomes zero.

For an $n$-type branching process there will be $n^2$ offspring distributions. 
Correspondingly there will be $n^2$ offspring means, which can be arranged into a matrix called the \textit{offspring mean matrix} $\boldsymbol{\Lambda}$. 
The criticality of the multi-type branching process is determined by the largest eigenvalue of $\mathbf{\Lambda}$.  
The process will always extinct if the largest eigenvalue of $\mathbf{\Lambda}$ is less than or equal to one \cite{KB:04,TE:89}.

Although the branching process does not directly represent any of the physics or mechanisms of the outage propagation, after it is validated it can be used to predict the total number of outages. 
The parameters of the branching process can be estimated from a much smaller data set, and then predictions of the total number of outages can be made based on the estimated parameters. 
The ability to do this via the branching process with much less data is a significant advantage that enables practical applications.

\section{Estimating Multi-Type Branching \\ Process Parameters} \label{estimating}

The simulation of OPA and its variants \cite{OPA1,OPA2,OPA3,OPA4,ACOPA,ACOPA1,slow} can naturally produce outages in generations. 
Each iteration of the ``main loop'' of the simulation produces another generation.
A total of $M$ cascades are simulated to produce nonnegative integer data that can be arranged as
\begin{table}[H]
\renewcommand{\arraystretch}{1.8}
\label{data}
\centering
\small
\begin{tabular}{cccc}
& generation\,0 & generation\,1 & $\cdots$ \\
\textrm{cascade 1} & $(Z^{1,1}_{0},\cdots,Z^{1,n}_{0})$ & $(Z^{1,1}_{1},\cdots,Z^{1,n}_{1})$ & $\cdots$ \\
\textrm{cascade 2} & $(Z^{2,1}_{0},\cdots,Z^{2,n}_{0})$ & $(Z^{2,1}_{1},\cdots,Z^{2,n}_{1})$ & $\cdots$ \\
$\vdots$ & $\vdots$ & $\vdots$ & $\vdots$ \\
\textrm{cascade} $M$ & $(Z^{M,1}_{0},\cdots,Z^{M,n}_{0})$ & $(Z^{M,1}_{1},\cdots,Z^{M,n}_{1})$ & $\cdots$ \\
\end{tabular}
\end{table}
\noindent
where $Z^{m,t}_{g}$ is the number of type $t$ outages in generation $g$ of cascade number $m$, and $n$ is the number of types of outages.
Each cascade has a nonzero number of outages in generation zero for at least one type of outages and 
each type of outage should have a nonzero number of outages at least for one generation. 
The shortest cascades stop in generation one, but some cascades will continue for several generations before terminating. 
Note that continuous data such as the load shed need to be first discretized by the method in \cite{bp13}.

Here, we explain how to estimate the offspring mean matrix and the empirical joint distribution of the initial outages from the simulated data. 
Note that we do not need all of $M$ cascades but only $M_u \le M$ cascades to perform the estimation. 
We will specially discuss the number of cascades needed to obtain a good estimate in Section \ref{num needed}.

\subsection{Estimating Offspring Mean Matrix} \label{estimate mean}

For $n$-type branching processes where $n\geq 2$ the offspring mean $\lambda$ will be generalized to the offspring mean matrix $\boldsymbol{\Lambda}$.
Different from the branching processes with only one type, for which the criticality is directly determined by
the offspring mean $\lambda$, the criticality of multi-type branching processes is determined by the largest eigenvalue of $\mathbf{\Lambda}$, which is denoted by $\rho$.
If $\rho \le 1$, the multi-type branching process will always extinct.
If $\rho > 1$, the multi-type branching process will extinct with a probability $0\le q <1$ \cite{TE:89}.

The largest eigenvalue $\rho$ of the mean matrix can be estimated as the total number of all types of children 
divided by the total number of all types of parents by directly using the simulated cascades and ignoring the types \cite{Guttorp}
\begin{equation} \label{rho}
\hat{\rho}=\frac{\sum\limits_{m=1}^{M_u}\sum\limits_{g=1}^{\infty}\sum\limits_{t=1}^{n}Z_g^{m,t}}
{\sum\limits_{m=1}^{M_u}\sum\limits_{g=0}^{\infty}\sum\limits_{t=1}^{n}Z_g^{m,t}}.
\end{equation}

When the number of type $j$ children to type $i$ parents $S^{(i,j)}$ and the total number of type $i$ parents $S^{(i)}$ are observed, $\lambda_{ij}$ (the expected number of type $j$ children generated by one type $i$ parent) can be estimated by a maximum likelihood estimator that is the total number of type $j$ children produced by type $i$ parents divided by the total number of type $i$ parents \cite{estimator}
\begin{equation} \label{mean}
\hat{\lambda}_{ij}=\frac{S^{(i,j)}}{S^{(i)}},
\end{equation}
where $S^{(i,j)}$ and $S^{(i)}$ can be described by using the simulated cascades as
\begin{align} \label{yij}
S^{(i,j)}&=\sum\limits_{m=1}^{M_u}\sum\limits_{g=1}^{\infty}Z_g^{m,i \rightarrow j} \\
S^{(i)}&=\sum\limits_{m=1}^{M_u}\sum\limits_{g=0}^{\infty}Z_g^{m,i},
\end{align}
where $Z_g^{m,i \rightarrow j}$ is the number of type $j$ offspring generated by type $i$ parents
in generation $g$ of cascade $m$.

However, it is usually impossible to have so detailed information.
For cascading blackouts, it is difficult to determine the exact number of type $j$ outages that are produced by type $i$ outages, due to too many mechanisms in cascading. 
In other words, $Z_g^{m,i \rightarrow j}$ in (\ref{yij}) cannot be determined and thus $Y^{(i,j)}$ cannot be decided and the mean matrix cannot be estimated.

To solve this problem, we apply the Expectation Maximization (EM) algorithm \cite{em}, 
which fits the problem well as a method for finding maximum likelihood estimates of parameters in statistical models where the model depends on unobserved latent variables. 
Besides, we assume the offspring distributions of branching processes are Poisson. 
There are general arguments suggesting that the choice of a Poisson offspring distribution is appropriate \cite{Kim12,bp13} since offspring outages being selected from a large number of possible outages have very small probability and are approximately independent. 
The EM algorithm mainly contains two steps, which are E-step and M-step.
For the estimation of the offspring mean matrix of an $n$-type branching process, the EM algorithm can be formulated as follows.

\begin{enumerate} \renewcommand{\labelitemi}{$\bullet$}

\item \textbf{Initialization:} Set initial guess of mean matrix as $\hat{\mathbf{\Lambda}}^{(0)}$. \\
Since for cascading blackouts the outages will always die out, we have $0\le \lambda_{ij} \le 1$. 
Based on this all elements of the initial mean matrix are set to be 0.5, which is the mid point of the possible range.

\vspace{0.1cm}
\item \textbf{E-step:} Estimate $S^{(i,j)(k+1)}$ based on $\hat{\mathbf{\Lambda}}^{(k)}$. \\
Under the assumption that the offspring distributions are all Poisson,
for generation $g\ge 1$ of cascade $m$, the number of type $j$ offspring produced by
type $t=1,\ldots,n$ parents follows Poisson distribution
\begin{align}
Z_g^{m,t\rightarrow j} & \sim \textrm{Pois}(Z_{g-1}^{m,t}\,\hat{\lambda}_{tj}^{(k)}).
\end{align}

Thus the number of type $j$ offspring in generation $g\ge 1$ of cascade $m$ produced by type $i$ parents in generation $g-1$ of the same cascade is:
\begin{align}
Z_g^{m,i\rightarrow j}=Z_g^{m,j}\,\frac{Z_{g-1}^{m,i}\hat{\lambda}_{ij}^{(k)}}{\sum\limits_{t=1}^{n}Z_{g-1}^{m,t}\hat{\lambda}_{tj}^{(k)}}.
\end{align}

After obtaining $Z_g^{m,i\rightarrow j}$ for all generations $g \ge 1$ of cascades $m=1,\ldots,M_u$ we are finally able to calculate $S^{(i,j)}$ by using (\ref{yij}).

\vspace{0.1cm}
\item \textbf{M-step:} Estimate $\hat{\mathbf{\Lambda}}^{(k+1)}$ based on $S^{(i,j)(k+1)}$.  \\
After obtaining $S^{(i,j)(k+1)}$ the updated mean matrix $\hat{\mathbf{\Lambda}}^{(k+1)}$ can be estimated with
the estimator given in (\ref{mean}).

\vspace{0.1cm}
\item \textbf{End:} Iterate the E-step and M-step until
\begin{equation} \label{end}
\max\limits_{i,j\in \{1,\cdots,n\}} |\hat{\lambda}_{ij}^{(k+1)}-\hat{\lambda}_{ij}^{(k)}|<\epsilon,
\end{equation}
where $\epsilon$ is the tolerance that is used to control the accuracy and $\hat{\lambda}_{ij}^{(k+1)}$ is the final estimate of $\lambda_{ij}$.

\end{enumerate}

\subsection{Estimating the Joint Distribution of Initial Outages}

For a $n$-type branching process, the empirical joint probability distribution of the number of 
initial outages $(Z_0^1,\cdots,Z_0^n)$ can be obtained as
\begin{align} \label{initial}
&d^{\textrm{emp}}_{Z_0}(z_0^1,\cdots,z_0^n)=P(Z_0^1=z_0^1,\cdots,Z_0^n=z_0^n) \notag \\
=&\frac{1}{M_u}\sum\limits_{m=1}^{M_u}I[Z^{m,1}_{0}=z_0^1,\cdots,Z_0^{m,n}=z_0^n],
\end{align}
where the notation $I[\textrm{event}]$ is the indicator function that evaluates to one when the event happens and evaluates to zero when the event does not happen.

\section{Estimating the Joint Probability Distribution of Total Outages} \label{total pdf}

Since we are most interested in the statistics of the total outages produced by the cascades, 
here we will discuss how to estimate the joint distribution of $n$ types of blackout size by using 
the estimated offspring mean matrix and the joint distribution of initial outages in Section \ref{estimating}.

\subsection{Estimation for an $n$-Type Branching Process} \label{n case}

The probability generating function for the type $i$ individual of an $n$-type branching process is
\begin{align}\label{gf}
f_i(s_1,\cdots,s_n)=\sum_{u1,\cdots,u_n=0}^\infty p_i&(u_1,\cdots,u_n)s_1^{u_1}\cdots s_n^{u_n},
\end{align}
where $p_i(u_1,\cdots,u_n)$ is the probability that a type $i$ individual generates
$u_1$ type 1, $\cdots$, $u_n$ type $n$ individuals. If we assume that the offspring distributions for various types of outages are all Poisson, as in Section \ref{estimating}, (\ref{gf}) can be easily written after the offspring mean matrix $\hat{\mathbf{\Lambda}}$ is estimated by the method in Section \ref{estimate mean}.

According to \cite{TE:89} and \cite{good}, the probability generating function, $w_i(s_1,\cdots,s_n)$, of the total number of various types of individuals in all generations, starting with one individual of type $i$, can be given by
\begin{equation}
w_i=s_i f_i(w_1,\cdots,w_n),\qquad i=1,\ldots,n.
\end{equation}

When the branching process starts with more than one type of individuals, the total number of various types
can be determined by using the Lagrange-Good inversion in \cite{good}, in which the following theorem is given.

\textit{Theorem 1:} If the $n$-type random branching process starts with $r_1$ individuals of type 1, $r_2$ of type 2, etc., 
then the probability that the whole process will have precisely $m_1$ of type 1, $m_2$ of type 2, etc., is equal to
the coefficient of
\begin{equation}
s_1^{m_1-r_1}\cdots s_n^{m_n-r_n} \notag
\end{equation}
in
\begin{equation}\label{det}
f_1^{m_1}\cdots f_n^{m_n}\biggr|\biggr|\delta_\mu^\nu-\frac{s_\mu}{f_\mu}\frac{\partial f_\mu}{\partial s_\nu}\biggr|\biggr|,
\end{equation}
where $||a_\mu^\nu||$ denotes the determinant of the $n\times n$ matrix whose entry is
$a_\mu^\nu (\mu,\nu=1,\ldots,n)$ and $\delta_\mu^\nu$ is Kronecker's delta
($=1$ if $\mu=\nu$, otherwise $=0$).
We denote the coefficient of $s_1^{m_1-r_1}\cdots s_n^{m_n-r_n}$ as
$c(r_1,\cdots,r_n;m_1,\cdots,m_n)$.

Given the joint probability distribution of initial sizes $P(Z_0^1,\cdots,Z_0^n)$
and the generating functions in (\ref{gf}), the formula for calculating
the joint probability distribution $d^{\textrm{est}}(y_1,\cdots,y_n)$ of the total number of various types
$(Y_{\infty}^1,\cdots,Y_{\infty}^n)$ can then be written as
\begin{align} \label{total}
&d^{\textrm{est}}_{Y_\infty}(y_1,\cdots,y_n)= P(Y_{\infty}^1=y_1,\cdots,Y_{\infty}^n=y_n) \notag \\
=&\sum\limits_{\substack{z_0^1,\cdots,z_0^n=0 \\ z_0^1+\cdots+z_0^n\neq 0}}^{z_0^1=y_1,\cdots,z_0^n=y_n} \biggr[P(Z_0^1=z_0^1,\cdots,Z_0^n=z_0^n)\LargerCdot \notag \\
&\qquad\qquad\qquad\qquad\quad c(z_0^1,\cdots,z_0^n;y_1,\cdots,y_n)\biggr].
\end{align}

\subsection{An Example for a Two-Type Branching Process}

Here, we take the joint probability distribution estimation of a two-type branching process as an example to better illustrate the proposed method.
The empirical joint probability distribution of the number of initial outages $(Z_0^1,Z_0^2)$ can be obtained 
by (\ref{initial}). 
As in Section \ref{estimating}, we assume that the offspring distributions for various types of outages are all Poisson. 
Then the probability generating functions for a two-type branching process can be written as
\begin{equation} \label{f1}
f_1(s_1,s_2)=\sum_{u1=u2=0}^{\infty}\frac{\lambda_{11}^{u_1}\lambda_{12}^{u_2}\,e^{-\lambda_{11}-\lambda_{12}}}{u_1!\,u_2!}s_1^{u_1}s_2^{u_2}
\end{equation}
\begin{equation} \label{f2}
f_2(s_1,s_2)=\sum_{u1=u2=0}^{\infty}\frac{\lambda_{21}^{u_1}\lambda_{22}^{u_2}\,e^{-\lambda_{21}-\lambda_{22}}}{u_1!\,u_2!}s_1^{u_1}s_2^{u_2},
\end{equation}
where the parameters $\lambda_{11}$, $\lambda_{12}$, $\lambda_{21}$, and $\lambda_{22}$ can be estimated by the method in Section \ref{estimate mean}.

In (\ref{det}) the $n\times n$ matrix whose determinant needs to be evaluated is actually
\begin{displaymath}
\renewcommand{\arraystretch}{2.4}
\left[ \begin{array}{cc}
\mathlarger{1-\frac{s_1}{f_1}\frac{\partial f_1}{\partial s_1}} & \mathlarger{-\frac{s_1}{f_1}\frac{\partial f_1}{\partial s_2}} \\
\mathlarger{-\frac{s_2}{f_2}\frac{\partial f_2}{\partial s_1}} & \mathlarger{1-\frac{s_2}{f_2}\frac{\partial f_2}{\partial s_2}}
\end{array} \right].
\end{displaymath}

The joint probability distribution of the two-type branching process 
can be obtained by evaluating (\ref{total}) with elementary algebra.
Since the coefficients in (\ref{f1}) and (\ref{f2}) will decrease very fast with the increase of the order of $s_1$ and $s_2$,
we can use a few terms to approximate the generating functions to reduce the calculation burden while guaranteeing accurate enough results.
Furthermore, the probability obtained by (\ref{total}) will also decrease with the increase of
$y_1$ and $y_2$. We do not need to calculate the negligible probability for too large blackout size.
Specifically, we can only calculate the joint probability for
\begin{equation} \label{y1}
y_1=z_0^1,\ldots,z_0^1+\tau_1
\end{equation}
and
\begin{equation} \label{y2}
y_2=z_0^2,\ldots,z_0^2+\tau_2,
\end{equation}
where $\tau_1$ and $\tau_2$ are integers properly chosen for a tradeoff of calculation burden and accuracy. 
Too large $\tau_1$ or $\tau_2$ will lead to unnecessary calculation for blackout sizes with negligible probability. 
Too small $\tau_1$ or $\tau_2$ will result in loss of accuracy by neglecting blackout sizes with not so small probability.


\subsection{Validation} \label{validate}

In Section \ref{n case} we propose a method to estimate the joint distribution of $n$ types of blackout size $(Y_\infty^1,\cdots,Y_\infty^n)$, which is denoted by $d^{\textrm{est}}_{Y_\infty}(y_1,\cdots,y_n)$. Here, we validate it by comparing it with the empirically obtained joint distribution $d^{\textrm{emp}}_{Y_\infty}(y_1,\cdots,y_n)$, which can be calculated by
\begin{align} \label{empi}
d^{\textrm{emp}}_{Y_{\infty}}(y_1,\cdots,y_n)=&P(Y_{\infty}^1=y_1,\cdots,Y_{\infty}^n=y_n) \notag \\
=&\frac{N(Y_{\infty}^1=y_1,\cdots,Y_{\infty}^n=y_n)}{M},
\end{align}
where $N(Y_{\infty}^1=y_1,\cdots,Y_{\infty}^n=y_n)$ is the number of cascades for which there are $y_1$ type 1 outages, $\cdots$, $y_n$ type $n$ outages.

Specifically, 
\begin{enumerate}
	\item \textbf{Joint entropy}: We compare them by the joint entropy, which can be defined for $n$ random variables $(Y_\infty^1,\cdots,Y_\infty^n)$ as
	\begin{align} \label{entropy}
	&H(Y_\infty^1,\cdots,Y_\infty^n) \notag \\
	=&-\sum\limits_{y_1}\cdots\sum\limits_{y_n}P(y_1,\cdots,y_n)\log_2[P(y_1,\cdots,y_n)],
	\end{align}
	where $P(y_1,\cdots,y_n)\log_2[P(y_1,\cdots,y_n)]$ is defined to be 0 if $P(y_1,\cdots,y_n)=0$.
	
	The joint entropy for the estimated and the empirical joint distribution can be respectively denoted by $H^{est}$ and $H^{emp}$. Then the estimated joint distribution can be validated by checking if $H^{\textrm{est}}/H^{\textrm{emp}}$ is close to 1.0.
	
	\item \textbf{Marginal distribution}: The marginal distribution for each type of outages can also be calculated after estimating the joint distribution of the total outages, 	which can be compared with the empirical marginal distribution directly calculated from the simulated cascades in order to validate the estimated joint distribution.
	
	\item \textbf{Conditional largest possible total outages (CLO)}: We can also calculate the conditional largest possible total outage of one type of blackout size when the total outage of the other types of blackout size are known. For example, for a two-type branching process, for $i,j\in \{1,2\}$ and $i \ne j$, given the total outage of one type of blackout size $y_i$ we can get the total outage of another type of blackout size $y_j$ that satisfies
	\begin{equation} \label{CLOeq}
	P(Y_{\infty}^j \le y_j|Y_{\infty}^i=y_i) = p_{\textrm{conf}}
	\end{equation}
	where 
	\begin{align}
	&P(Y_{\infty}^j \le y_j|Y_{\infty}^i=y_i) \notag \\
	=&\sum\limits_{k=0}^{y_j} \frac{P(Y_\infty^i=y_i,Y_\infty^j=k)}{\sum\limits_{l=0}^{\infty}P(Y_\infty^i=y_i,Y_\infty^j=l)},
	\end{align}
	$p_{\textrm{conf}}$ is the confidence level close to $1.0$ and $P(A|B)$ is the conditional probability of event A given B. 
	If we know that the total outage of type $i$ is $y_i$, from the joint distribution we know that the total outage of type $j$
	will not exceed $y_j$ with a high probability $p_{\textrm{conf}}$.
	
	We can calculate the $y_j$ from either the empirical joint distribution or the estimated joint distribution by branching process
	and compare them to check if the $y_j$ from the estimated joint distribution is close to that from the empirical joint distribution. 
\end{enumerate}

\section{Number of Cascades Needed} \label{num needed}

In the above sections we assume there are a total of $M$ cascades and in section \ref{estimating}
we use $M_u$ of them to estimate the offspring mean matrix and the joint distribution of initial outages. 
But two questions remain unanswered, which are how many cascades we need to empirically obtain a reliable joint distribution of total outages 
and how many cascades we need to get a reliable estimate of the offspring mean matrix and joint distribution of initial outages which can further guarantee that the estimated joint distribution of total outages is close enough to the reliable empirical joint distribution. 
Here, we discuss these questions by a similar method in \cite{interaction} and determine the lower bounds $M^{\min}$ and $M_u^{\min}$ respectively for $M$ and $M_u$.

\subsection{Determining Lower Bound for $M$} \label{Mmin}

More cascades tend to contain more information about the property of cascading failures of a system. 
The added information brought from the added cascades will make the joint entropy of the joint distribution empirically obtained from the cascades increase. 
However, the amount of information will not always grow with the increase of the number of cascades but will saturate after the number of cascades is greater than some number $M^{\min}$, 
which can be determined by gradually increasing the number of cascades, recording the corresponding joint entropy of the empirical joint distribution, and finding the smallest number of cascades that can lead to the saturated joint entropy (amount of information).

Assume there are a total of $N_M$ different $M$'s ranging from a very small number to a very large number, 
which are denoted by $M_i,\,i=1,2,\ldots,N_M$.
The joint entropy of the joint distribution of total outages obtained from $M_i$ cascades is denoted by $H^{\textrm{emp}}(M_i)$.

For $i=1,\ldots,N_M-2$ we define 
\begin{align}
\sigma_i&=\sigma(H^{\textrm{emp}}_i),
\end{align}
where $H^{\textrm{emp}}_i=\mathlarger{[}H^{\textrm{emp}}(M_i) \; \cdots \; H^{\textrm{emp}}(M_{N_M})\mathlarger{]}$ 
and $\sigma(\cdot)$ is the standard deviation of a vector.
The $\sigma_i$ for $i=N_M-1$ and $i=N_M$ are not calculated since we want to calculate the standard deviation for at least three data points.
Very small and slightly fluctuating $\sigma_i$ indicates that the joint entropy begins to saturate after $M_i$.
Specifically, the $M_i$ corresponding to $\sigma_i\le \epsilon_\sigma$ 
is identified as $M^{\min}$, where $\epsilon_\sigma$ is a small real number.

The $M^{\min}$ original cascades can guarantee that the accuracy on statistical values of interest is good and thus can provide a reference joint distribution of the total outages.

\subsection{Determining Lower Bound for $M_u$} \label{deter Mu}

When we only want to obtain good enough estimate of the joint distribution of total sizes, 
we do not need as many as $M^{\min}$ cascades but only $M_u^{\min}$ cascades to make sure that the information extracted from $M_u^{\min}$ cascades by 
the branching process can capture the general properties of the cascading failures. Here, we propose a method to determine $M_u^{\min}$.

Since both $H^{\textrm{emp}}$ and $H^{\textrm{est}}$ vary with $M_u$, we denote them by $H^{\textrm{emp}}(M_u)$ and $H^{\textrm{est}}(M_u)$. 
$H^{\textrm{emp}}(M_u)$ can be directly obtained from the cascades by (\ref{empi}) and (\ref{entropy}) and $H^{\textrm{est}}(M_u)$ can be calculated by (\ref{total}) and (\ref{entropy}).

When $M_u$ is not large enough, it is expected that there will be a big mismatch between $H^{\textrm{emp}}(M_u)$ and $H^{\textrm{est}}(M_u)$, 
indicating that the estimated joint distribution from the branching process cannot well capture the property of the joint distribution of the cascades. 
But with the increase of $M_u$ more information will be obtained and thus the mismatch will gradually decrease and finally stabilization. 
In order to indicate the stabilization, we define 
\begin{equation}
R(M_u)=\frac{|H^{\textrm{est}}(M_u)-H^{\textrm{emp}}(M_u)|}{H^{\textrm{emp}}(M_u)},
\end{equation}
start from a small integer $M_u^0$ and increase it gradually by $\Delta M$ each time, and calculate the standard deviation of $R(M_u)$ for the latest three data points by
\begin{equation}
\tilde{\sigma}_i=\sigma(R_i), \;\; i\ge 2,
\end{equation}
where $i$ denotes the latest data point and 
\begin{equation}
R_i=\mathlarger{[}R(M_u^{i-2}) \; R(M_u^{i-1}) \; R(M_u^{i})\mathlarger{]}.
\end{equation}
Then $M_u^{\min}$ is determined as the smallest value that satisfies $\tilde{\sigma}_i \le \epsilon_{H}$ 
where $\epsilon_{H}$ is used to determine the tolerance for stabilization.

By decreasing $\Delta M$ we can increase the accuracy of the obtained $M_u^{\min}$. 
But smaller $\Delta M$ will increase the times of calculating the joint distribution by branching processes. 
When more types of outages are considered, greater $M_u^{\min}$ will be needed, 
in which case larger $\Delta M$ can be chosen to avoid too many times of calculating the joint distribution.

Note that there is an implicit assumption in this section that all the cascades are generated from the same cascading failure model or at least from 
similar models. If the cascades come from very different cascading failure models or are generated by very different mechanisms, it might be possible that the proposed methods in this section are difficult to converge or stabilize with the increase of the number of cascades.

\section{Results}\label{simulation}

Here we present results of the branching process parameters computed from simulated cascades 
and the joint distributions of outages predicted from these parameters.
The cascading outage dataset is produced by the open-loop AC OPA simulation \cite{ACOPA,ACOPA1} on the IEEE 118-bus test system, 
which is standard except that the line flow limits are determined with the same method in \cite{bp13}.
The probability for the initial line outage is $p_{0}=0.0001$ and the load variability $\gamma=1.67$, which are the same as \cite{Ian10} and \cite{bp13}.

To test the multi-type branching process model, the simulation is run so as to produce $M=50000$ cascades with a nonzero number of line outages at the base case load level. 
In each generation the number of line outages and the number of isolated buses are counted and the continuously varying amounts of the load shed are discretized as described in \cite{bp13} 
to produce integer multiples of the chosen discretization unit.

\subsection{Number of Cascades Needed}

The method in Section \ref{num needed} is used to determine $M^{\min}$ and $M_u^{\min}$. 
For determining $M^{\min}$, we choose $N_M=50$ and the data points are linearly scaled. 
The $\epsilon_\sigma$ is chosen as $0.002$. In order to determine $M_u^{\min}$, we choose $M_u^0$, $\epsilon_{H}$, 
and $\Delta M$ in Section \ref{deter Mu} as $100$, $0.002$, 
and 100 for one type of outages and $1000$, $0.002$, and $500$ for multiple types of outages, since the $M_u^{\min}$ for multiple outages case 
is expected to be greater and we need to limit the calculation burden. 
The determined $M^{\min}$ and $M_u^{\min}$ for different types of outages are listed in Table \ref{num}.
The $M_u^{\min}$ used for estimation is significantly smaller than $M^{\min}$, thus helping greatly improve the efficiency.

\begin{table}[!t]
\renewcommand{\arraystretch}{1.5}
\footnotesize
\caption{Number of Cascades Needed}
\label{num}
\centering
\footnotesize
\begin{tabular}{cccc}
\hline
\hline
No. of types & Type & $M^{\min}$ & $M_u^{\min}$ \\
\hline
1 & line outage & 18000 & 1400 \\
1 & load shed & 36000 & 1900 \\
1 & isolated bus & 33000 & 900 \\
2 & \tabincell{c}{line outage and \\ load shed} & 39000 & 6500 \\
2 & \tabincell{c}{line outage and \\ isolated bus} & 37000 & 5500 \\
\hline
\hline
\end{tabular}
\end{table}

\subsection{Parameters of Branching Processes}

The $\epsilon$ in (\ref{end}) is chosen as $0.01$. The EM algorithm that is used to estimate the offspring mean matrix of 
the multi-type branching processes can quickly converge. The number of iterations $N^{\textrm{ite}}$ is listed in Table \ref{iteration}.

\begin{table}[!t]
\renewcommand{\arraystretch}{1.6}
\footnotesize
\caption{Number of Iterations of EM Algorithm}
\label{iteration}
\centering
\begin{tabular}{ccc}
\hline
\hline
Type & $M_u$ & $N^{\textrm{ite}}$ \\
\hline
line outage and load shed & \tabincell{c}{39000 \\ 6500} & \tabincell{c}{7 \\ 7}  \\
\hline
line outage and isolated bus & \tabincell{c}{37000 \\ 5500} & \tabincell{c}{4 \\ 4}  \\
\hline
\hline
\end{tabular}
\end{table}

The estimated branching process parameters are listed in Table \ref{para1}, where $\hat{\lambda}$ is the offspring mean for one type of outages 
estimated by the method in \cite{bp13}. 
It is seen that the estimated largest eigenvalue of the offspring mean matrix $\hat{\rho}$ is greater than the estimated offspring means 
for only considering one type of outages, indicating that the system is closer to criticality when we simultaneously consider two types of outages. 
This is because different types of outages, such as line outage and the load shed, can mutually influence each other, 
thus aggregating the propagation of cascading. 
In this case, only considering one type of outages will underestimate the extent of outage propagation. 

The $\hat{\lambda}_{12}$ in $\hat{\boldsymbol{\Lambda}}$ is the estimated expected discretized number of the load shed when one line is tripped while $\hat{\lambda}_{21}$ is the estimated expected number of line outages when one discretization unit of load is shed. 
From the offspring mean matrix $\hat{\boldsymbol{\Lambda}}$ we can see that line outages tend to have a greater influence on the load shed and isolated buses 
but the influence of the load shed or isolated buses on line outages is relatively weak. 
This is reasonable since in real blackouts it is more possible for line tripping to cause the load shed or isolated buses. 
Sometimes line outages directly cause the load shed or isolated buses, for example, the simplest case occurs when a load is fed from a radial line. 

Also note that there is some mismatch between the largest eigenvalue of the offspring mean matrix $\hat{\rho}$ estimated from (\ref{rho}) 
and that calculated from the estimated offspring mean matrix $\hat{\boldsymbol{\Lambda}}$ by the EM algorithm. 
The estimator in (\ref{rho}) is the maximum likelihood estimator of the largest eigenvalue of the offspring mean matrix \cite{Guttorp}, which does not need to make any assumption about the offspring distribution. By contrast, in order to estimate the offspring mean matrix we have to assume a specific offspring distribution, such as the Poisson distribution used in this paper. 
As mentioned in Section \ref{estimate mean}, there are general arguments suggesting that the choice of a Poisson offspring distribution is appropriate, which will also be further validated in the following sections. However, the offspring distribution is only approximately Poisson but not necessarily exactly Poisson. 
Numerical simulation of multi-type branching processes with Poisson offspring distributions shows that the estimated $\hat{\rho}$ and the largest eigenvalue of the estimated $\hat{\boldsymbol{\Lambda}}$ do agree with each other. Therefore, the largest eigenvalue estimated from (\ref{rho}) without any assumption of the offspring distribution is expected to be more reliable and the closeness of the system to criticality should thus be determined based on the estimated $\hat{\rho}$ from (\ref{rho}).

\begin{table}[!t]
\renewcommand{\arraystretch}{1.6}
\footnotesize
\caption{Estimated Parameters of Branching Processes by (\ref{rho}) and the EM Algorithm Using $M^{\min}$ Cascades }
\label{para1}
\centering
\begin{tabular}{ccccc}
\\[-15pt]
\hline
\hline
Type  & $\hat{\lambda}$ & $\hat{\rho}$ & $\hat{\boldsymbol{\Lambda}}$ \\
\hline
line outage & 0.45 & -- & -- \\
load shed & 0.48 & -- & -- \\
isolated bus & 0.14 & -- & -- \\
\tabincell{c}{line outage and \\ load shed} & -- & 0.55 & $
\renewcommand{\arraystretch}{1.2}
\left[ \begin{array}{cc}
0.45 & 0.42 \\
0.0018 & 0.029
\end{array} \right] $  \\ [4ex]
\tabincell{c}{line outage and  \\ isolated bus} & -- & 0.60 & $
\renewcommand{\arraystretch}{1.2}
\left[ \begin{array}{cc}
0.45 & 0.40 \\
6.0\times 10^{-5} & 0.0049
\end{array} \right] $ \\ [4ex]
\hline
\hline
\end{tabular}
\end{table}

The estimated parameters for branching processes by only using $M_u^{\min}$ cascades are listed in Table \ref{para2}, which are 
very close to those estimated by using $M^{\min}$ cascades, indicating that $M_u^{\min}$ cascades are enough to get good estimate.

\begin{table}[!t]
\renewcommand{\arraystretch}{1.6}
\footnotesize
\caption{Estimated Parameters of Branching Processes by (\ref{rho}) and the EM Algorithm Using $M_u^{\min}$ Cascades}
\label{para2}
\centering
\begin{tabular}{ccccc}
\\[-15pt]
\hline
\hline
Type & $\hat{\lambda}$ & $\hat{\rho}$ & $\hat{\boldsymbol{\Lambda}}$ \\
\hline
line outage & 0.45 & -- & -- \\
load shed & 0.49 & -- & -- \\
isolated bus & 0.15 & -- & -- \\
\tabincell{c}{line outage and \\ load shed} & -- & 0.56 & $
\renewcommand{\arraystretch}{1.2}
\left[ \begin{array}{cc}
0.45 & 0.43 \\
0.0020 & 0.027
\end{array} \right] $  \\ [4ex]
\tabincell{c}{line outage and  \\ isolated bus} & -- & 0.61 & $
\renewcommand{\arraystretch}{1.2}
\left[ \begin{array}{cc}
0.45 & 0.39 \\
5.5\times 10^{-5} & 0.0040
\end{array} \right] $  \\ [4ex]
\hline
\hline
\end{tabular}
\end{table}

\subsection{Estimating Joint Distribution of Total Outages} 

In (\ref{f1}) and (\ref{f2}), the highest orders for both $s_1$ and $s_2$ are chosen as $4$. 
In (\ref{y1}) and (\ref{y2}), $\tau_1$ and $\tau_2$ are chosen based on the number of initial outages 
from the samples of cascades and the tradeoff between calculation burden and accuracy. 
For line outages and the load shed, $\tau_1$ and $\tau_2$ are chosen as $12$ and $9$, respectively. 
For line outages and isolated buses, $\tau_1$ and $\tau_2$ are chosen as $12$ and $18$, respectively. 

It has been shown for one-type branching process that it is much more time efficient to estimate the parameters of a branching process from a shorter simulation run and then predict the distribution of total outages by branching process than it is to run much longer simulation in order to accumulate enough cascades to empirically estimate the distribution \cite{Ian10,Kim12,bp13}.
Here, we estimate the joint distribution of total outages with multi-type branching process by using $M_u^{\min}\ll M^{\min}$ cascades and compare it with 
the empirical joint distribution obtained from $M^{\min}$ cascades. 

To quantitatively compare the empirical and estimated joint distributions, the joint entropy is calculated and 
listed in Table \ref{entro}. It is seen that the joint entropy of the estimated joint distributions is reasonably close 
to that of the empirical joint distributions. Also, the joint entropy of the distributions for two types of outages is significantly greater than 
that for one type of outages, meaning that we can get new information by jointly analyze two types of outages.

\begin{table}[!t]
\renewcommand{\arraystretch}{1.6}
\footnotesize
\caption{Joint Entropy of Distributions}
\label{entro}
\centering
\begin{tabular}{cccc}
\\[-15pt]
\hline
\hline
Type & $M_u$ & $H^{\textrm{emp}}$ & $H^{\textrm{est}}$ \\
\hline
line outage & \tabincell{c}{18000 \\ 1400} & \tabincell{c}{3.50 \\ 3.48} & \tabincell{c}{3.91 \\ 3.92} \\
\hline
load shed & \tabincell{c}{36000 \\ 1900} & \tabincell{c}{3.52 \\ 3.53} & \tabincell{c}{3.56 \\ 3.57} \\
\hline
isolated bus & \tabincell{c}{33000 \\ 900} & \tabincell{c}{2.63 \\ 2.59} & \tabincell{c}{2.64 \\ 2.61} \\
\hline
line outage and load shed & \tabincell{c}{39000 \\ 6500} & \tabincell{c}{6.99 \\ 6.94} & \tabincell{c}{7.08 \\ 7.06} \\
\hline
line outage and isolated bus & \tabincell{c}{37000 \\ 5500} & \tabincell{c}{5.33 \\ 5.30} & \tabincell{c}{6.45 \\ 6.44} \\
\hline
\hline
\end{tabular}
\end{table}

\begin{figure}[!t]
\centering
\includegraphics[width=2.8in]{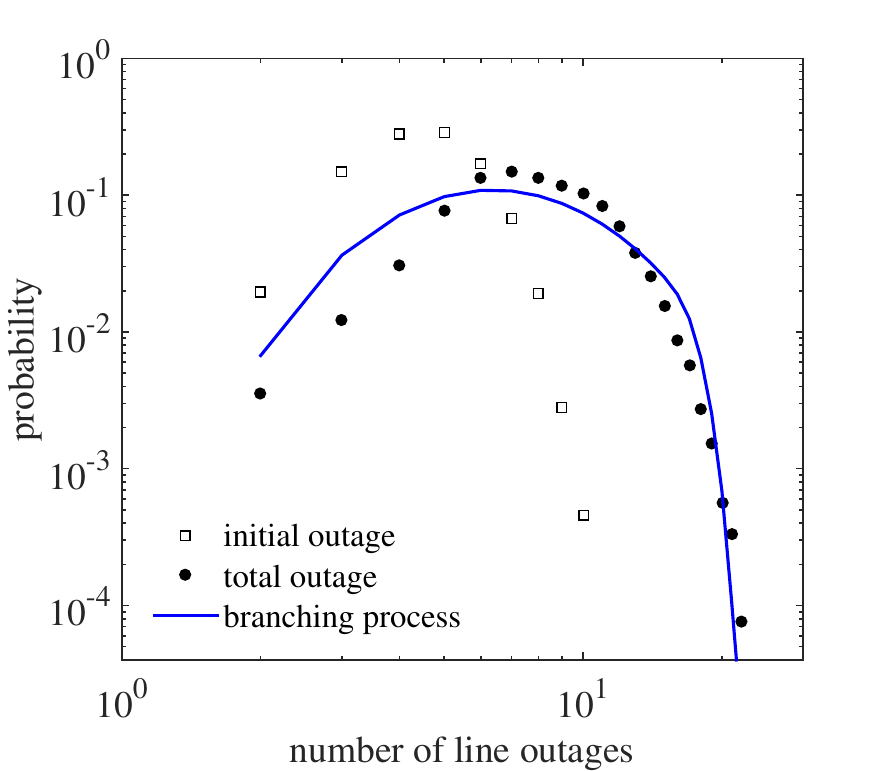}
\caption{Estimated marginal probability distribution of the number of line outages by using $M_u^{\min}=6500$ cascades when line outages and the load shed are considered. 
Dots indicate total outages and squares indicate initial outages; both distributions are empirically obtained from $M^{\min}=39000$ simulated cascades. 
The solid line indicates the distribution of total outages predicted with the multi-type branching process.}
\label{line1}
\end{figure}

\begin{figure}[!t]
\centering
\includegraphics[width=2.8in]{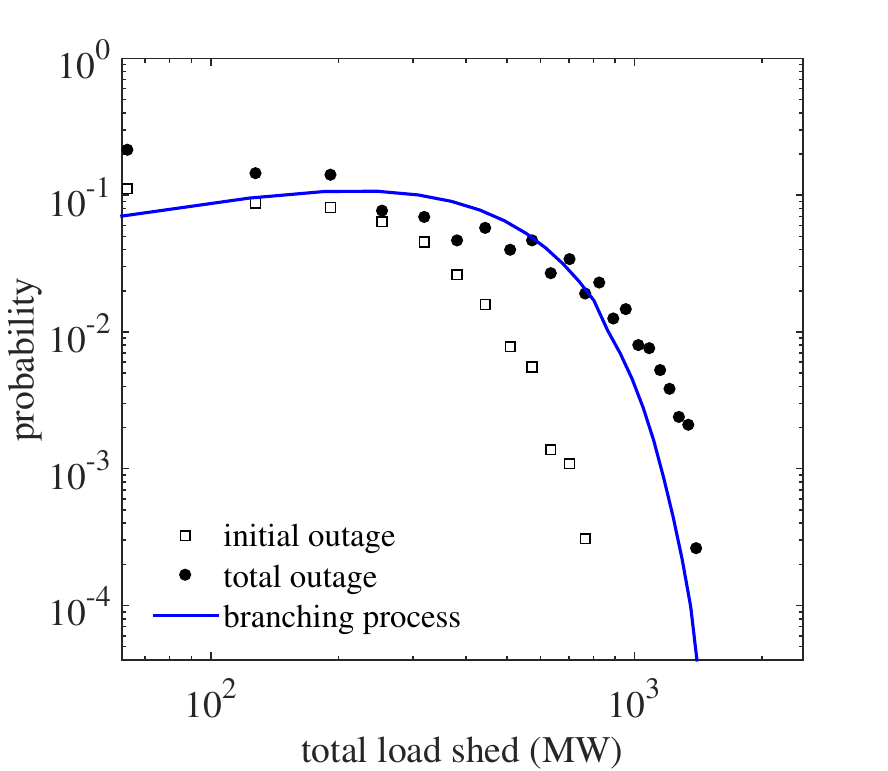}
\caption{Estimated marginal probability distribution of the load shed by using $M_u^{\min}=6500$ cascades.}
\label{load1}
\end{figure}

\begin{figure}[!t]
\centering
\includegraphics[width=2.8in]{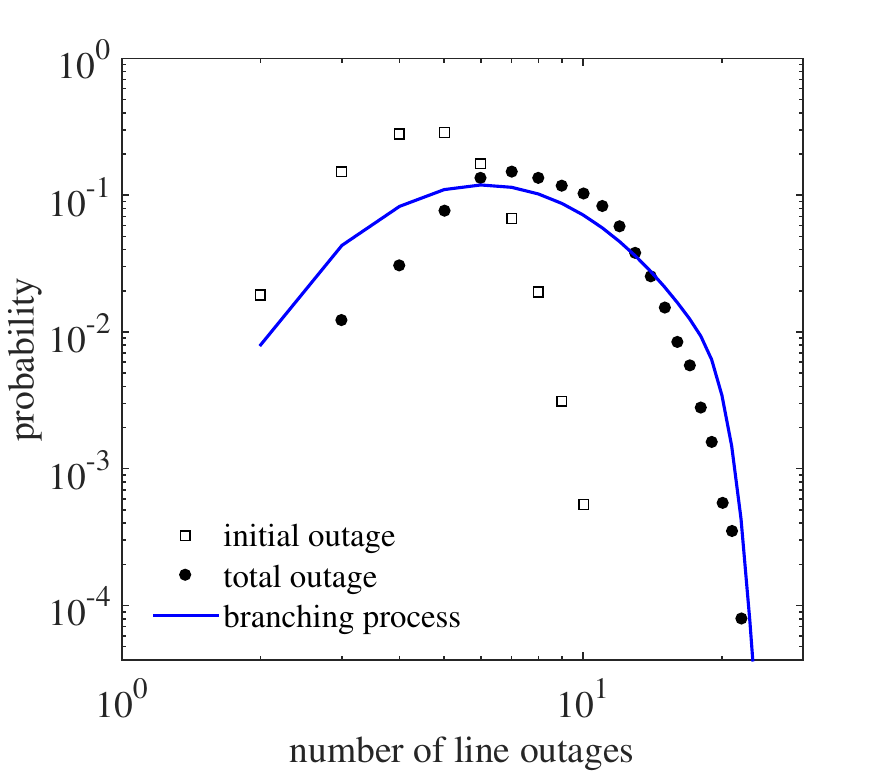}
\caption{Estimated marginal probability distribution of the number of line outages by using $M_u^{\min}=5500$ cascades when the line outages and isolated buses are considered. 
Dots indicate total outages and squares indicate initial outages; both distributions are empirically obtained from $M^{\min}=37000$ simulated cascades. 
The solid line indicates the distribution of total outages predicted with the multi-type branching process. }
\label{line2}
\end{figure}

\begin{figure}[!t]
\centering
\includegraphics[width=2.8in]{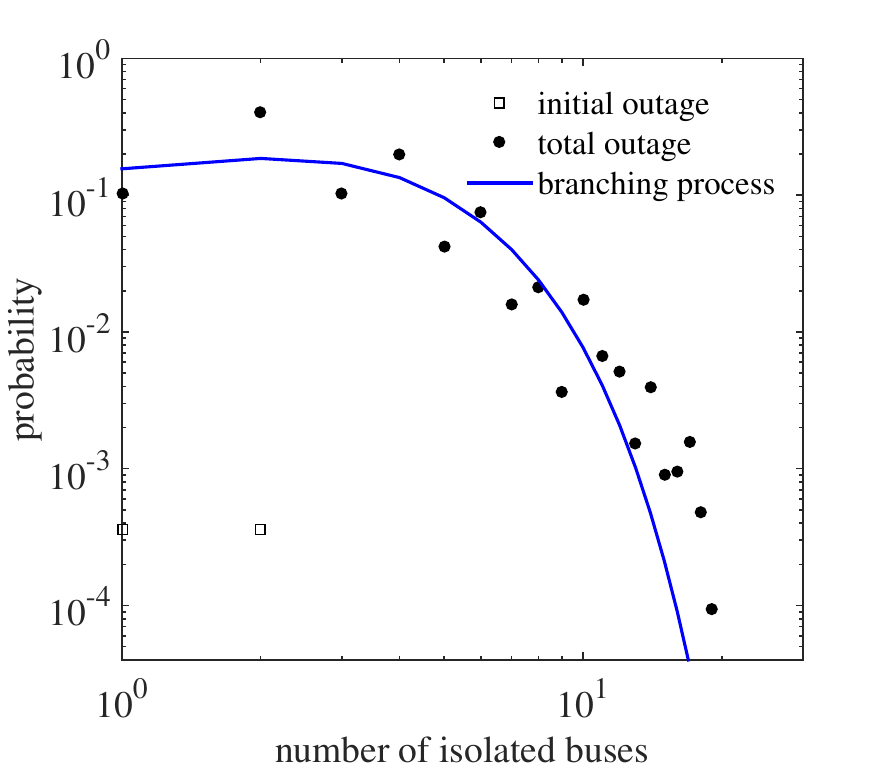}
\caption{Estimated marginal probability distribution of the number of isolated buses by using $M_u^{\min}=5500$ cascades.}
\label{bus2}
\end{figure}

After estimating the joint distributions, the marginal distributions for each type of outage can also be calculated. 
In Figs. \ref{line1} and \ref{load1} we show the marginal distribution of line outages and the load shed for the two-type branching process of 
line outages and the load shed. The empirical marginal distributions of total outages (dots) and initial outages (squares) calculated from $M^{\min}=39000$ 
are shown, as well as a solid line indicating the total outages predicted by the multi-type branching process from $M_u^{\min}=6500$ cascades. 
The branching process data is also discrete, but is shown as a line for ease of comparison. 
It is seen that the branching process prediction with $M_u^{\min}=6500$ cascades matches the marginal distribution 
empirically obtained by using $M^{\min}=39000$ cascades very well. 
Similar results for the marginal distribution of the line outages and isolated buses for the two-type branching process of line outages and isolated buses are shown in Figs. \ref{line2} and \ref{bus2}, for which $M^{\min}=37000$ and $M_u^{\min}=5500$.

\begin{figure}[!t]
\centering
\includegraphics[width=2.8in]{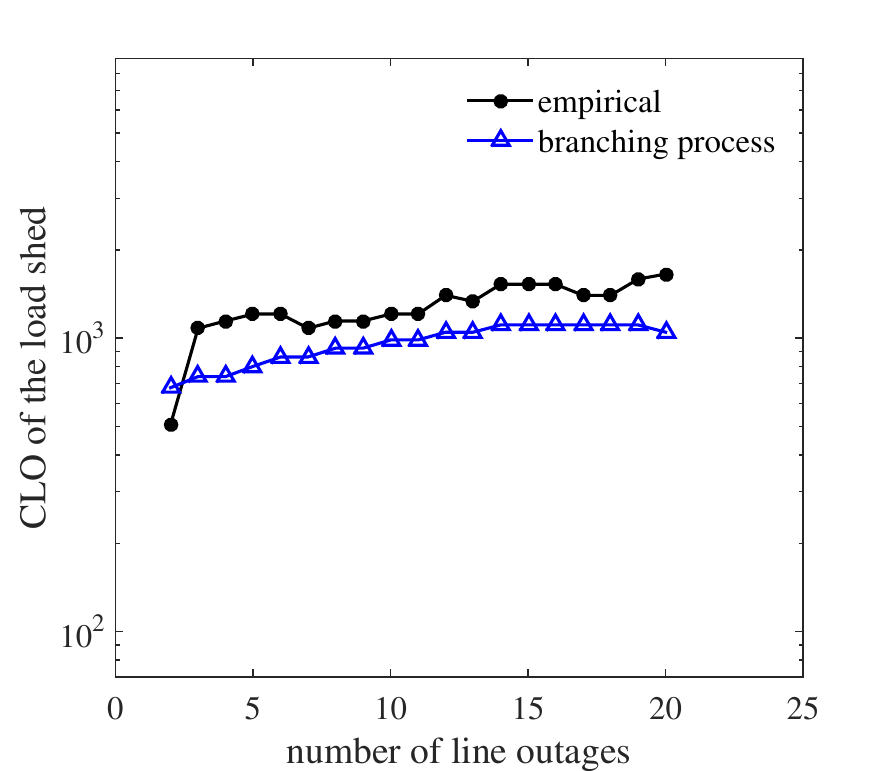}
\caption{Estimated CLO for the load shed when the total number of line outages is known.}
\label{clo1}
\end{figure}

\begin{figure}[!t]
\centering
\includegraphics[width=2.8in]{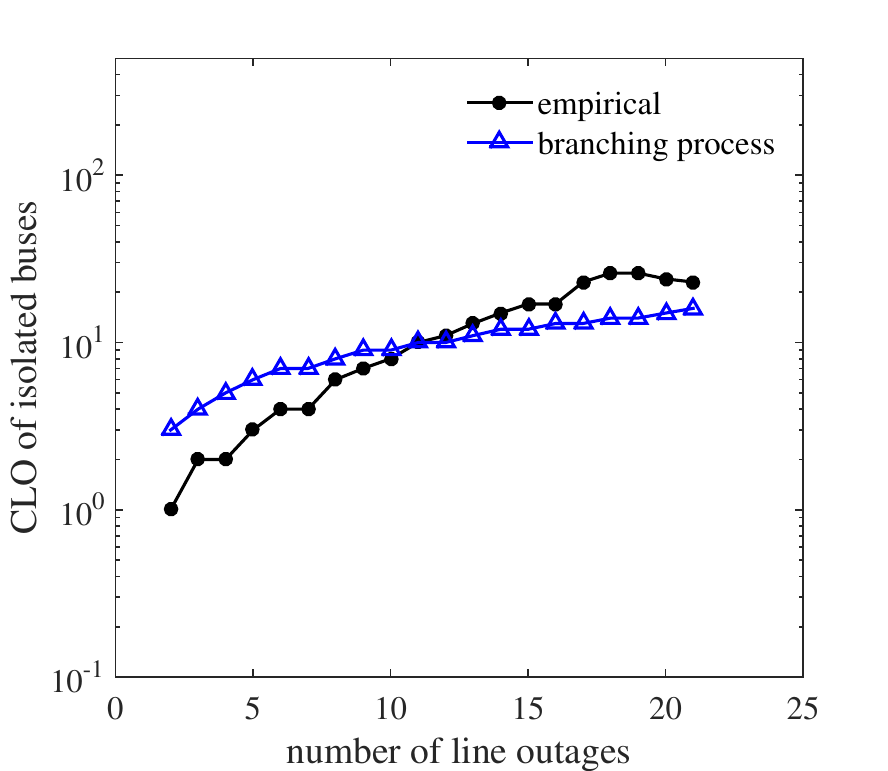}
\caption{Estimated CLO for the isolated buses when the total number of line outages is known.}
\label{clo2}
\end{figure}

The conditional largest possible total outages (CLO) defined in Section \ref{validate} when the total number of line outages is known can also be calculated from either the empirical joint distribution using $M^{\min}$ cascades or from the estimated joint distribution from branching process using $M_u^{\min} \ll M^{\min}$ cascades. In this paper the $p_{\textrm{conf}}$ in (\ref{CLOeq}) is chosen as $0.99$. 
The CLOs for the load shed and the isolated buses are, respectively, shown in Figs. \ref{clo1}--\ref{clo2}, which indicate that the CLO estimated by multi-type branching process using a much smaller number of cascades matches the empirically obtained CLO very well.

In order to further validate the proposed method for estimating the joint distribution of the total outages, 
we also perform a thorough cross validation. Specifically, randomly chosen $M_u^{\min}$ cascades are used to estimate the joint distribution by the multi-type branching process, 
which is compared with the joint distribution empirically obtained from another randomly chosen $M_u^{\min}$ cascades. 
The corresponding results, as shown in the Appendix, show that the model trained by some randomly chosen subset of data is accurate in describing other subsets.

\subsection{Predicting Joint Distribution from One Type of Outage}

To further demonstrate and validate the proposed multi-type branching process model, 
we estimate the joint distribution of the total sizes of two types of outages by only using the predetermined offspring mean matrix and 
the distribution of initial line outages, as follows.

\begin{enumerate}
	\item The offspring mean matrix $\hat{\boldsymbol{\Lambda}}$ is calculated offline from $M_u^{\min}$ cascades, as shown in Table \ref{para2}.
	\item To mimic online application, $M_u^{\min}$ cascades are randomly chosen from the $M^{\min} - M_u^{\min}$ cascades for test. 
	The empirical joint distribution of line outages and the load shed (isolate buses) is calculated as a reference. 
	\item We estimate the joint distribution of line outage and the load shed (isolated buses) by using the $\hat{\boldsymbol{\Lambda}}$ in step 1 and the distribution of initial line outages, 
	assuming there are no data about the load shed (isolated buses) for which initial outage is set to be zero with probability one.
	\item We compare the marginal distributions and the CLO calculated from the estimated and empirical joint distributions. 
\end{enumerate}

\begin{figure}[!t]
\centering
\includegraphics[width=2.8in]{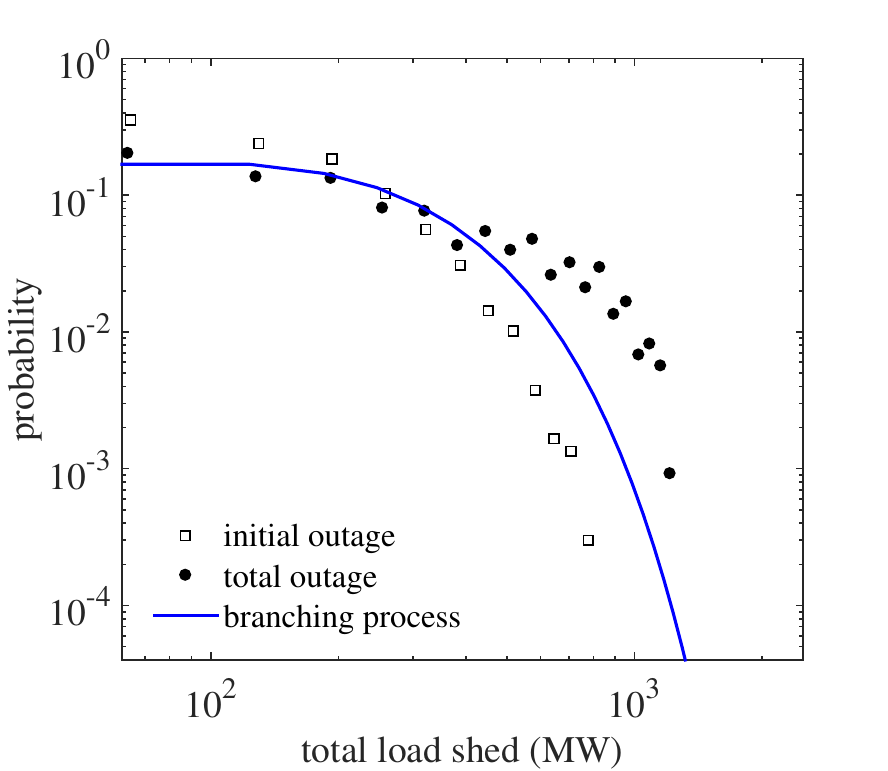}
\caption{Estimated marginal probability distribution of the load shed assuming there are no load shed data.}
\label{load3}
\end{figure}

\begin{figure}[!t]
\centering
\includegraphics[width=2.8in]{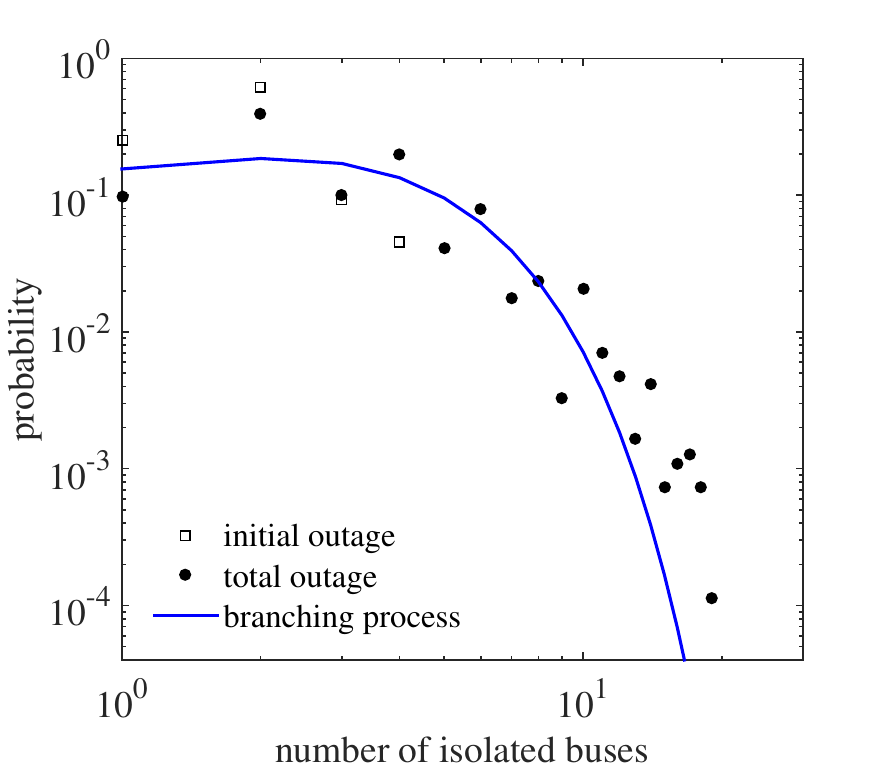}
\caption{Estimated marginal probability distribution of number of isolated buses assuming there are no isolated bus data.}
\label{bus4}
\end{figure}

The predicted marginal probability distributions of the load shed and isolated buses are shown in Figs. \ref{load3} and \ref{bus4}. 
The prediction is reasonably good even if we do not have the distribution of initial load shed or isolated buses. 
Also, from Fig. \ref{load3} it is seen that the prediction of the load shed is very good 
when the blackout size is small while the prediction when the blackout size is large is not as good. By contrast, the prediction of the number of isolated buses is good for both small and large blackout sizes. 
This is mainly because the initial outage of the load shed can be greater than zero with a nonnegligible probability 
and assuming the initial outage is zero with probability one can influence the accuracy of the prediction. 
However, the initial number of isolated buses is zero or one with a high probability ($86.21\%$ in this case) since in the initial stage the possibility that some buses are isolated from the major part of the system is very low, and thus assuming the initial number of isolated buses is zero with probability one does not obviously influence the prediction. 

\begin{figure}[!t]
\centering
\includegraphics[width=2.8in]{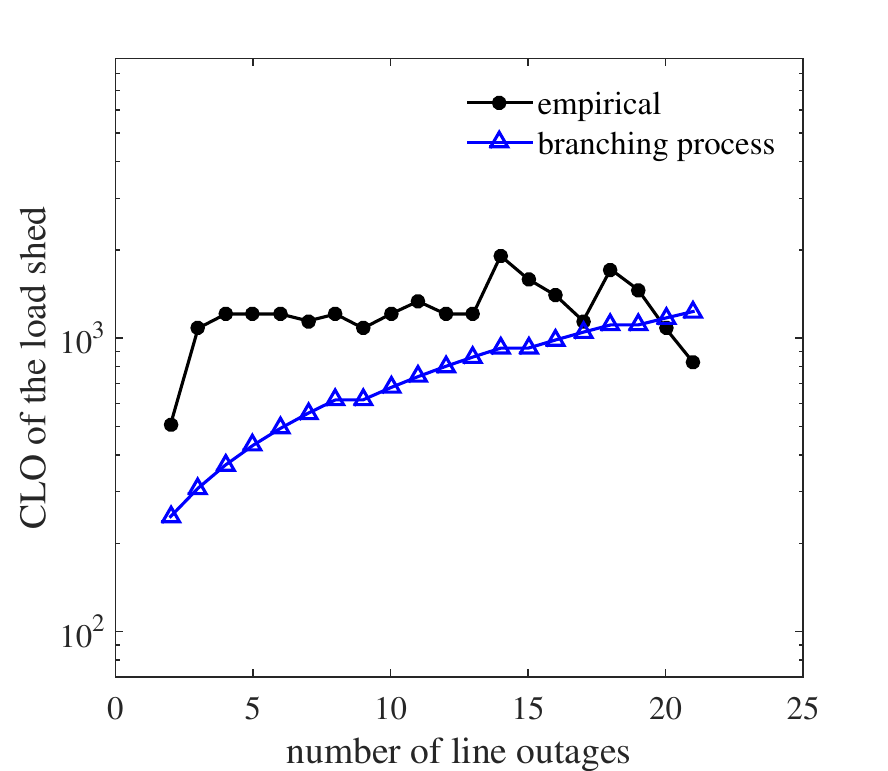}
\caption{Estimated CLO for the load shed when the number of line outages is known assuming there are no load shed data.}
\label{clo3}
\end{figure}

\begin{figure}[!t]
\centering
\includegraphics[width=2.8in]{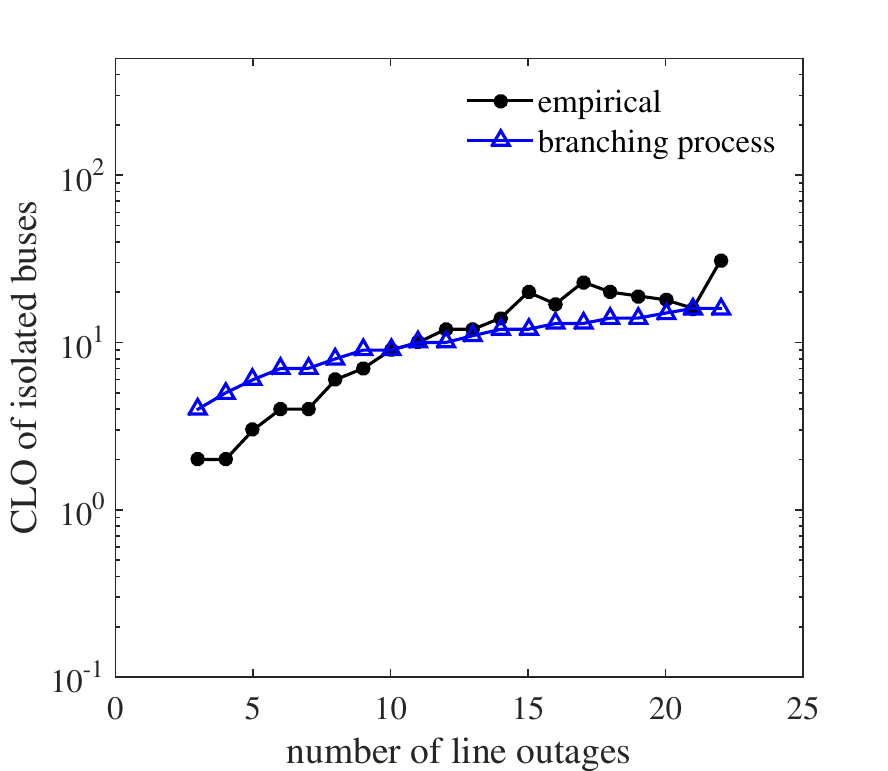}
\caption{Estimated CLO for the isolated buses when the number of line outages is known assuming there are no isolated bus data.}
\label{clo4}
\end{figure}

The empirically obtained and estimated CLOs of the load shed and the isolated buses when the number of line outages is known are shown in Figs. \ref{clo3}--\ref{clo4}, respectively. In both figures we use $M_u^{\min}$ cascades to get the empirical and estimated CLOs. 
The prediction of the CLO when there are no data for the load shed or isolated buses (especially the former one) is not as good as the case with those data (the prediction of the CLO for the isolated buses is better than that for the load shed for the same reason as that for the prediction of the marginal distribution discussed above).
However, the multi-type branching processes can generate useful and sometimes very accurate predictions for those outages whose data are unavailable, which can further provide important information for the operators when the system is under a cascading outage event or is in restoration. 
It is also seen that the estimated CLO from the branching process seems to be more statistically reliable than the empirically obtained CLO from the same number of cascades which can oscillate as the number of line outages increases. Comparing the empirically obtained CLOs in Fig. \ref{clo3} and Fig. \ref{clo1}, we can see that the oscillation in Fig. \ref{clo1} is not that obvious, mainly because it uses much more simulated cascades to obtain the empirical CLO.

\subsection{Estimating Propagation of Three Types of Outages}

In the above sections we only consider up to two types of outages, mainly because the calculation complexity for estimating the joint distribution for more than two types of outages can significantly increase. However, we can estimate the parameters of the three-type branching process, which can be used to better indicate the 
extent of the outage propagation. 
By using the method in Section \ref{Mmin}, we determine $M^{\min}=46000$ when considering line outages, the load shed, and isolated buses simultaneously. 
The EM algorithm for estimating the offspring mean matrix of the multi-type branching processes converges in six steps. 
The estimated largest eigenvalue of offspring mean matrix, the offspring mean matrix, and 
the joint entropy of the empirical joint distribution are listed in Table \ref{three}. 
It is seen that line outages tend to have a greater influence on the load shed and
isolated buses but the influence of the load shed or isolated buses on line outages is relatively weak.
The largest eigenvalue of the offspring mean matrix is greater than that for the two-type branching processes, 
indicating that the system is even closer to criticality when considering the mutual influence of three types of outages. 
Besides, the joint entropy is also greater compared with the two-type branching process, 
although the increase of joint entropy from two-type to three-type is not as high as that from one-type to two-type.

\begin{table}[!t]
\renewcommand{\arraystretch}{1.6}
\footnotesize
\caption{Estimated Parameters for Three Types of Outages}
\label{three}
\centering
\footnotesize
\begin{tabular}{cccc}
\hline
\hline
Type & $\hat{\rho}$ & $\hat{\Lambda}$ & $H^{\textrm{emp}}$ \\
\hline
\tabincell{c}{line outage, \\ load shed, and \\ isolated bus} & 0.64 & $
\renewcommand{\arraystretch}{1.0}
\left[ \begin{array}{c@{\hspace{0.4em}}c@{\hspace{0.4em}}c@{\hspace{0.4em}}}
0.44 & 0.39 & 0.39 \\
0.0035 & 0.024 & 0.018 \\
7.4\times 10^{-7} & 0.079 & 4.2\times 10^{-4}
\end{array} \right] $ & 8.54 \\ [4ex]
\hline
\hline
\end{tabular}
\end{table}

\section{Conclusion} \label{conclusion}

In this paper, the multi-type branching process is applied to statistically describe the propagation of line outages, the load shed, and isolated buses. 
The largest eigenvalues of the estimated offspring mean matrix for more than one type of outages are greater than the offspring means 
for one type of outages, indicating that the system is actually closer to criticality and only considering one type of outages will underestimate the extent of outage propagation. 
The joint distributions of two types of outages are efficiently estimated by the multi-type branching process with much smaller number of cascades than 
empirically estimating the joint distribution, which is a significant advantage since simulation time is a limiting factor when studying cascading blackouts.
The example studied suggests that the multi-type branching process can effectively predict the distribution of the load shed and isolated buses and their conditional largest possible total outages
even when there are no data for them. Finally, we demonstrate that a three-type branching process can provide joint analyses on line outages, the load shed, and isolated buses.


\appendix

\begin{figure}[!t]
\centering
\includegraphics[width=2.8in]{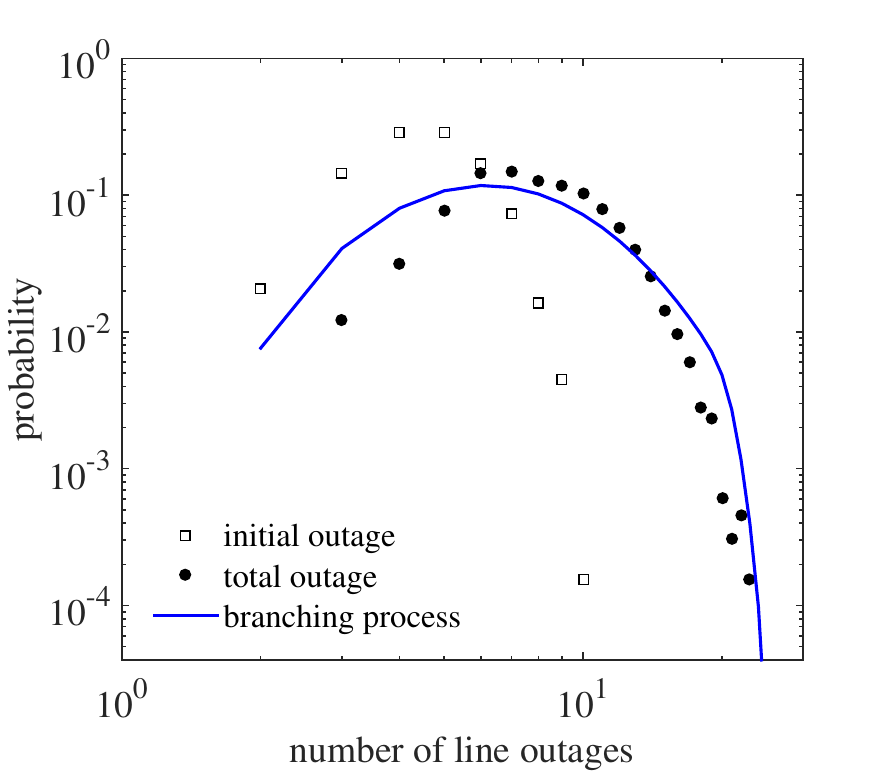}
\caption{Estimated marginal probability distribution of the number of line outages in cross validation when line outages and load shed are considered.}
\label{cross_line1}
\end{figure}

\begin{figure}[!t]
\centering
\includegraphics[width=2.8in]{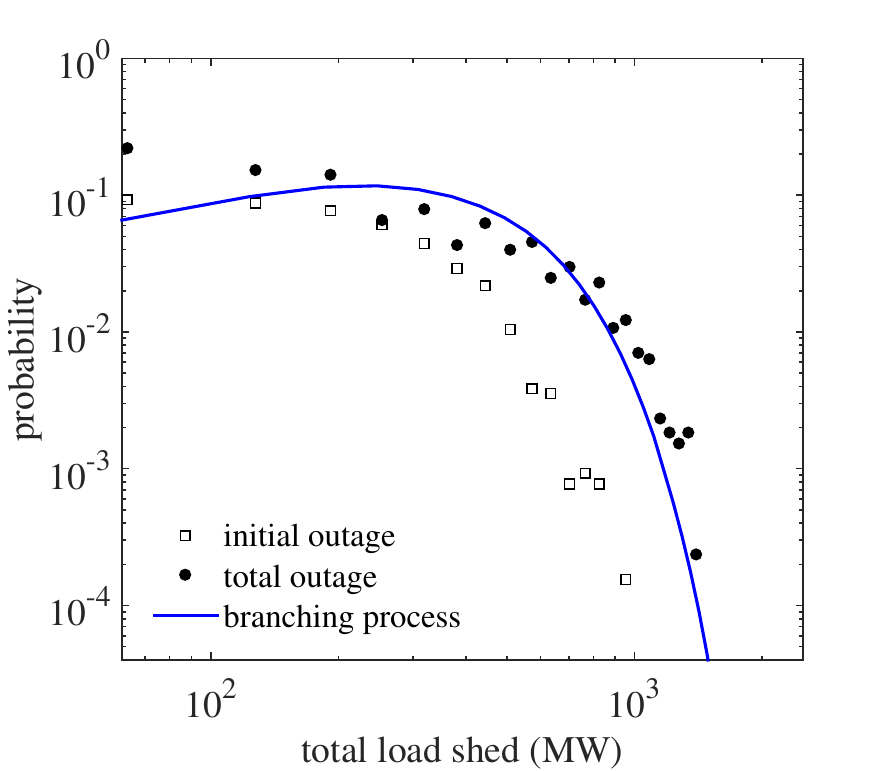}
\caption{Estimated marginal probability distribution of the load shed in cross validation.}
\label{cross_load1}
\end{figure}

\begin{figure}[!t]
\centering
\includegraphics[width=2.8in]{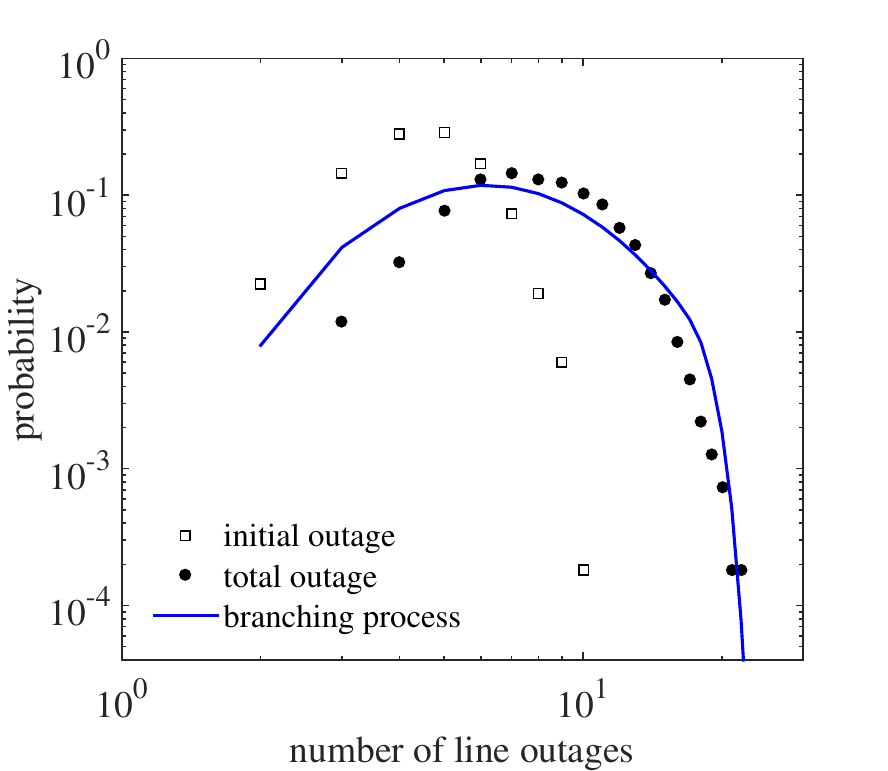}
\caption{Estimated marginal probability distribution of the number of line outages in cross validation when line outages and isolated buses are considered.}
\label{cross_line2}
\end{figure}

\begin{figure}[!t]
\centering
\includegraphics[width=2.8in]{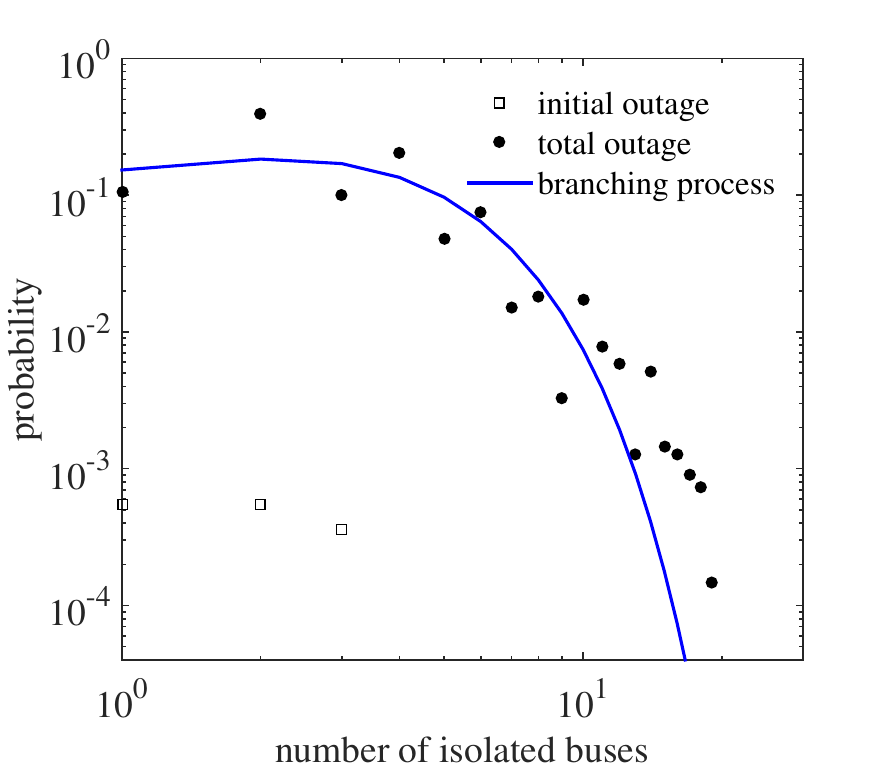}
\caption{Estimated marginal probability distribution of the number of isolated buses in cross validation.}
\label{cross_bus2}
\end{figure}

\begin{figure}[!t]
\centering
\includegraphics[width=2.8in]{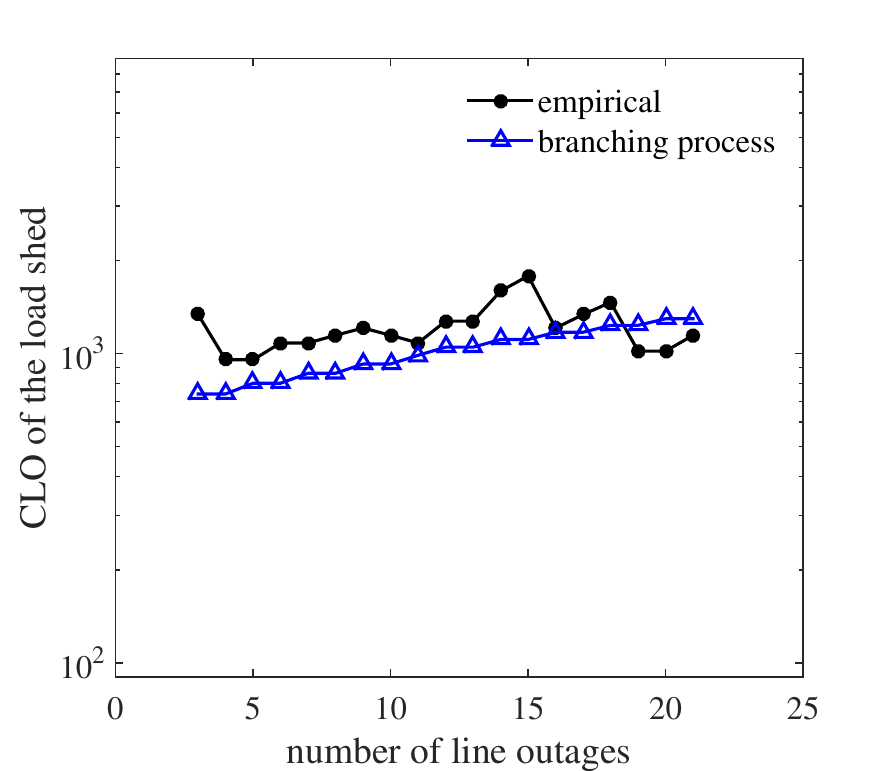}
\caption{Estimated CLO for the load shed in cross validation when the total number of line outages is known.}
\label{clo5}
\end{figure}

\begin{figure}[!t]
\centering
\includegraphics[width=2.8in]{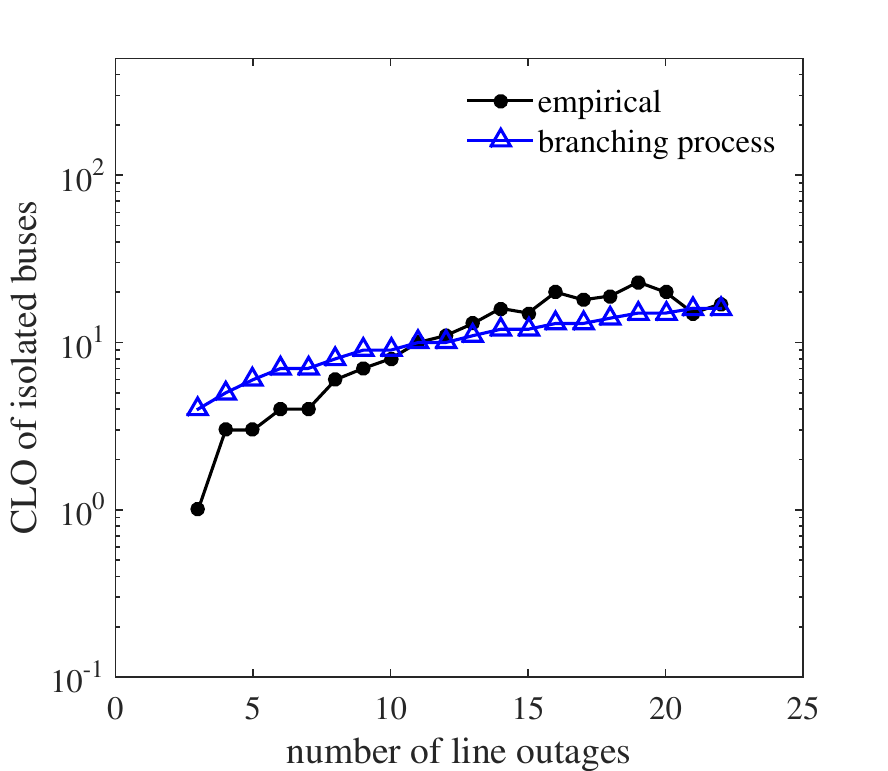}
\caption{Estimated CLO for the isolated buses in cross validation when the total number of line outages is known.}
\label{clo6}
\end{figure}

This appendix presents results for cross validation. Randomly chosen $M_u^{\min}$ cascades are used to estimate the joint distribution by the multi-type branching process, 
which is compared with the joint distribution empirically obtained from another randomly chosen $M_u^{\min}$ cascades. 
The joint entropy for the empirical and estimated joint distributions is listed in Table \ref{entro1}. 
The marginal distributions for each type of outages are shown in Figs. \ref{cross_line1}--\ref{cross_bus2}. 
The empirically obtained and estimated CLOs for the load shed and the isolated buses when the total number of line outages is known are shown in Figs. \ref{clo5}--\ref{clo6}. 
The results show that the branching process model trained by some randomly chosen subset of data is accurate in describing other subsets.

\begin{table}[H]
\renewcommand{\arraystretch}{1.6}
\footnotesize
\caption{Joint Entropy of Distributions in Cross Validation}
\label{entro1}
\centering
\begin{tabular}{cccc}
\hline
\hline
Type & $M_u^{\min}$ & $H^{\textrm{emp}}$ & $H^{\textrm{est}}$ \\
\hline
line outage and load shed & 6500 & 6.91 &  7.23 \\
\hline
line outage and isolated bus & 5500 & 5.33 & 6.38 \\
\hline
\hline
\end{tabular}
\end{table}

\ifCLASSOPTIONcaptionsoff
  \newpage
\fi

\begin{IEEEbiography} [{\includegraphics[width=1in,height=1.25in,clip,keepaspectratio]{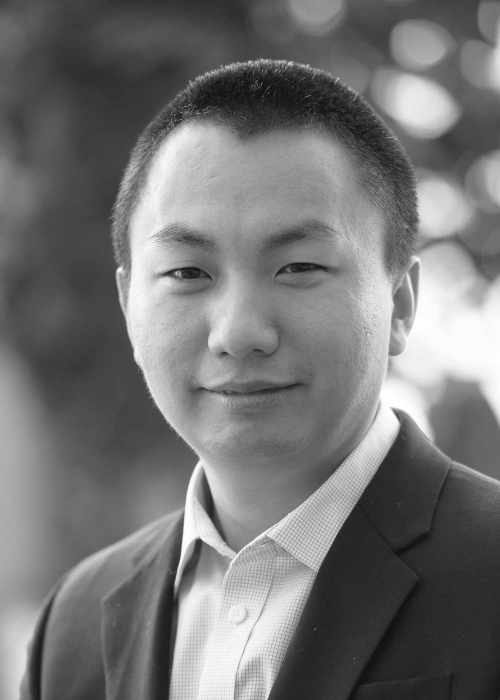}\vfill}]
	{Junjian Qi} (S'12--M'13)
	received the B.E. degree from Shandong University, Jinan, China, in 2008 and the Ph.D. degree Tsinghua University, Beijing, China, in 2013, both in electrical engineering.
	
	In February--August 2012 he was a Visiting Scholar at Iowa State University, Ames, IA, USA. During September 2013--January 2015 he was 
	a Research Associate at Department of Electrical Engineering and Computer Science, University of Tennessee, Knoxville, TN, USA. 
	Currently he is a Postdoctoral Appointee at the Energy Systems Division, Argonne National Laboratory, Argonne, IL, USA. 
	His research interests include cascading blackouts, power system dynamics, state estimation, synchrophasors, and cybersecurity.
\end{IEEEbiography}

\begin{IEEEbiography} [{\includegraphics[width=1in,height=1.25in,clip,keepaspectratio]{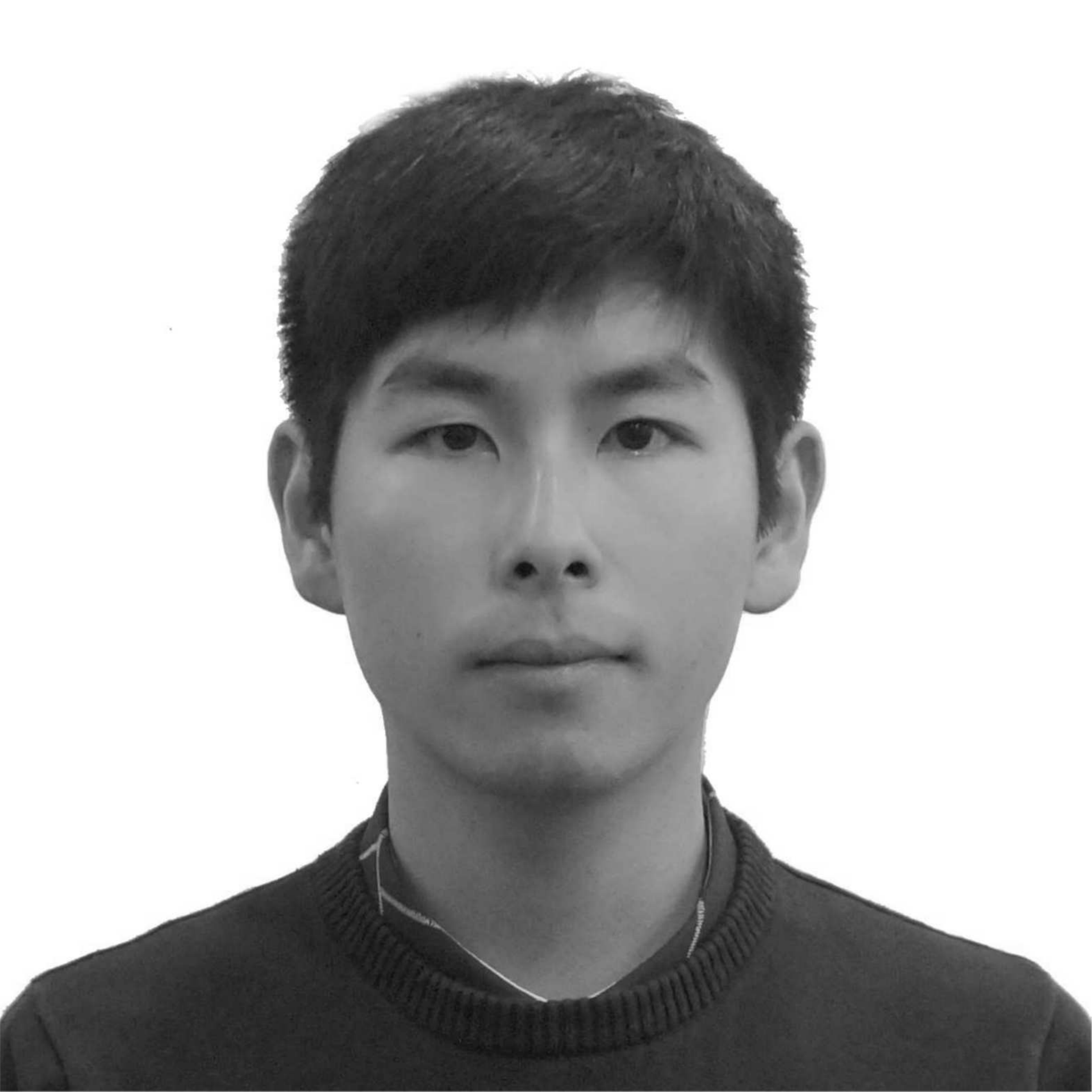}\vfill}]
	{Wenyun Ju} (S'15)
	received the B.E. degree in electrical information from Sichuan University, Chengdu, China in 2010, and M.Sc. degree in 
	electrical and electronic engineering from Huazhong University of Science and Technology, Wuhan, China in 2013. 
	
	Currently he is pursuing Ph.D. degree in the Department of Electrical Engineering and Computer Science, University of Tennessee, Knoxville, TN, USA.
	His research interests include cascading blackouts and vulnerability assessment of power grids.
	
\end{IEEEbiography}

\begin{IEEEbiography} [{\includegraphics[width=1in,height=1.25in,clip,keepaspectratio]{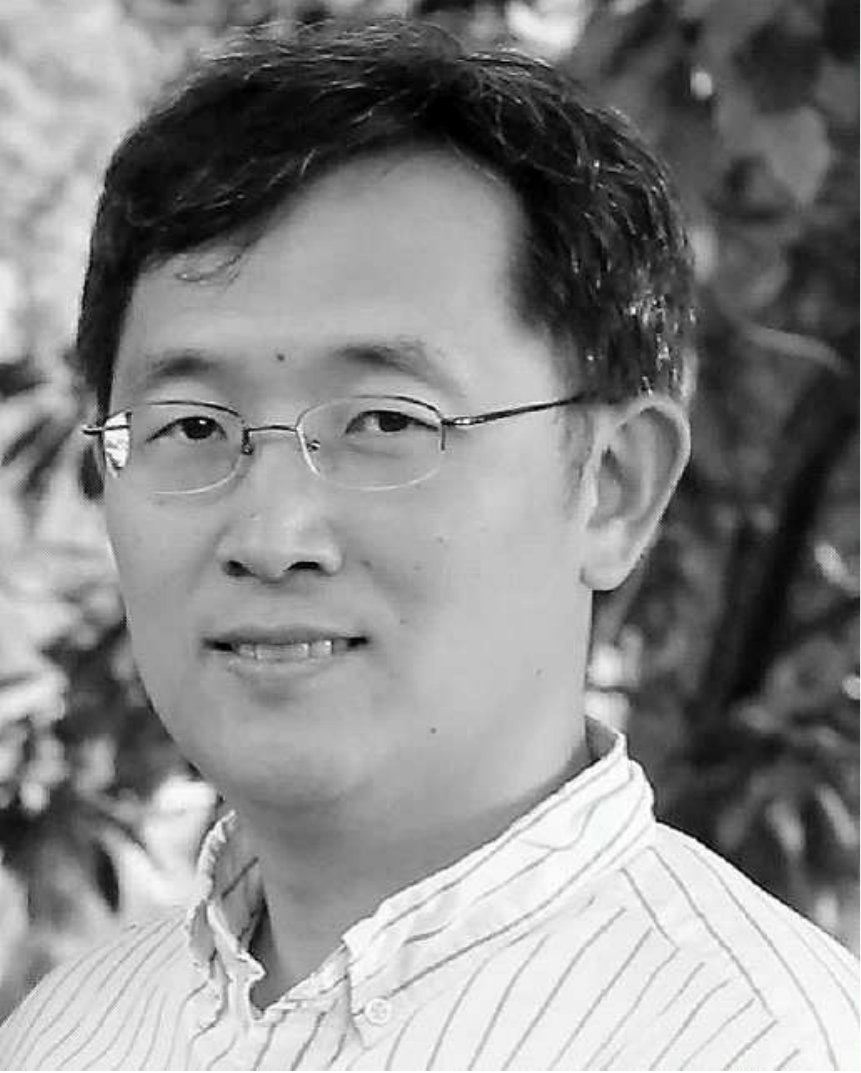}\vfill}]
	{Kai Sun} (M'06--SM'13)
	received the B.S. degree in automation in 1999 and the Ph.D. degree in control science and engineering in 2004 both from Tsinghua University, Beijing, China. 
	
	He is currently an assistant professor at the Department of Electrical Engineering and Computer Science, University of Tennessee, Knoxville, TN, USA. 
	He was a project man-ager in grid operations and planning at the EPRI, Palo Alto, CA from 2007 to 2012. 
	Dr. Sun is an editor of IEEE Transactions on Smart Grid and an associate editor of IET Generation, Transmission and Distribu-tion. 
	His research interests include power system dynamics, stability and control and complex systems.
\end{IEEEbiography}


\begin{thebibliography}{11}

\bibitem{usblackout}
U.S.-Canada Power System Outage Task Force, ``Final report on the August 14th blackout in the United States and Canada,'' Apr. 2004.

\bibitem{nerc}
NERC (North America Electric Reliability Council), ``1996 System Disturbances,'' (Available from NERC, Princeton Forrestal Village, 116--390 Village Boulevard, Princeton, New Jersey), 2002.

\bibitem{TF}
IEEE PES CAMS Task Force on Cascading Failure, ``Initial review of  methods for cascading failure analysis in electric power transmission systems,"
in \emph{Proc. IEEE Power and Energy Society General Meeting}, Pittsburgh PA USA, pp. 1--8, Jul. 2008.

\bibitem{Motter}
A.~E. Motter and Y. Lai, ``Cascade-based attacks on complex networks,'' \emph{Physical Review E}, vol. 66, no. 6, 065102(R), Dec. 2002.

\bibitem{zhao}
L. Zhao,  K. Park, and Y. Lai, ``Attack vulnerability of scale-free networks due to cascading breakdown,'' \emph{Physical Review E}, vol. 70, no. 3, 035101, Sept. 2004.

\bibitem{sandpile}
J. Qi and S. Pfenninger, ``Controlling the self-organizing dynamics in a sandpile model on complex networks by failure tolerance,'' \emph{EPL (Europhysics Letters)}, vol. 111, no. 3, 38006, Aug. 2015.

\bibitem{OPA1}
I. Dobson, B.~A. Carreras, and V.~E. Lynch, ``An initial model for complex dynamics in electric power system blackouts,'' \emph{34th Hawaii Intl. Conference on System Sciences}, HI, pp. 710--718, Jan. 2001.

\bibitem{OPA2}
B. A. Carreras, V. E. Lynch, I. Dobson, and D. E. Newman,
``Critical points and transitions in an electric power transmission model for cascading failure blackouts,"
{\it Chaos}, vol. 12, pp. 985-994, Dec. 2002.

\bibitem{OPA3}
H. Ren, I. Dobson, and B. A. Carreras,
``Long-term effect of the n-1 criterion on cascading line outages in an evolving power transmission grid,"
 {\it IEEE Trans. Power Syst.}, vol. 23, pp. 1217-1225, Aug. 2008.

\bibitem{OPA4}
B. A. Carreras, D. E. Newman, I. Dobson, and N. S. Degala,
``Validating OPA with WECC data," \emph{46th Hawaii Intl. Conference on System Sciences}, HI, Jan. 2013.

\bibitem{ACOPA}
S. Mei, Yadana, X. Weng, and A. Xue, ``Blackout model based on OPF and its self-organized criticality,'' \emph{Proceedings of the 25th Chinese Control Conference}, pp. 7--11, 2006.

\bibitem{ACOPA1}
S. Mei, Y, Ni. Weng, G. Wang, and S. Wu, ``A study of self-organized criticality of power system under cascading failures based on AC-OPA with voltage stability margin,'' \emph{IEEE Trans.  Power Systems}, vol. 23, pp. 1719--1726, Nov. 2008.

\bibitem{slow}
J. Qi, S. Mei, and F. Liu, ``Blackout model considering slow process,'' \emph{IEEE Trans. Power Syst.}, vol. 28, pp. 3274--3282, Aug. 2013.

\bibitem{kirschen}
M. A. Rios, D. S. Kirschen, D. Jawayeera, et al, ``Value of security: modeling time-dependent phenomena and weather conditions,'' \emph{IEEE Trans. Power Systems}, vol. 17, pp. 543--548, Aug. 2002.

\bibitem{thorp}
A. G. Phadke and J. S. Thorp, ``Expose hidden failures to prevent cascading outages,'' \emph{IEEE Comput. Appl. Power}, vol. 9, no. 3, pp. 20--23, Jul. 1996. 

\bibitem{chen}
J. Chen, J. S. Thorp, and I. Dobson. ``Cascading dynamics and mitigation assessment in power system disturbances via a hidden failure model,'' \emph{Int. J. Elect. Power Energy Syst.}, vol. 27, no. 4, pp. 318--326, May 2005. 

\bibitem{dynamic}
J. Song, E. Cotilla-Sanchez, G. Ghanavati, and P. H. Hines, ``Dynamic modeling of cascading failure in power systems,'' \emph{IEEE Trans. Power Syst.}, vol. 31, no. 3, pp. 2085--2095, May 2016.

\bibitem{interaction}
J. Qi, K. Sun, and S. Mei, ``An interaction model for simulation and mitigation of cascading failures," \emph{IEEE Trans. Power Syst.}, vol. 30, no. 2, pp. 804--819, Mar. 2015.

\bibitem{interaction1}
W. Ju, J. Qi, and K. Sun, ``Simulation and analysis of cascading failures on an NPCC power system test bed," \emph{IEEE Power and Energy Society General Meeting}, Denver CO, pp. 1--5, Jul. 2015.

\bibitem{linegraph}
P. D. Hines, I. Dobson, E. Cotilla-Sanchez, and M. Eppstein, ````Dual Graph" and ``Random Chemistry" methods for cascading failure analysis," \emph{46th Hawaii Intl. Conference on System Sciences}, HI, Jan. 2013.

\bibitem{Ian10}
I. Dobson, J.~Kim, and K.~R. Wierzbicki, ``Testing branching process estimators of cascading failure with data from a simulation of transmission line outages,'' \emph{Risk Analysis}, vol. 30, no. 4, pp. 650--662, Apr. 2010.

\bibitem{Kim12}
J.~Kim, K.~R. Wierzbicki, I. Dobson, and R.~C. Hardiman, ``Estimating propagation and distribution of load shed in simulations of cascading blackouts,'' \emph{IEEE Systems Journal}, vol. 6, no. 3, pp. 548-557, Sept. 2012.

\bibitem{bp13}
J. Qi, I. Dobson, and S. Mei, ``Towards estimating the statistics of simulated cascades of outages with branching processes,'' \emph{IEEE Trans. Power Syst.}, vol. 28, no. 3, pp. 3410--3419, Aug. 2013.

\bibitem{Ren08}
H. Ren and I. Dobson, ``Using transmission line outage data to estimate cascading failure propagation in an electric power system,'' \emph{IEEE Trans. Circuits and Systems Part \uppercase\expandafter{\romannumeral2}}, vol. 55, no. 9, pp. 927--931, Sept. 2008.

\bibitem{Ian12}
I. Dobson, ``Estimating the propagation and extent of cascading line outages from utility data with a branching process,"
{\sl IEEE Trans. Power Syst.},  vol. 27, no. 4, pp. 2146--2155, Nov. 2012.

\bibitem{KB:04}
K.~B. Athreya and P.~E. Ney, \emph{Branching Processes}, Dover NY, 2004.

\bibitem{TE:89}
T.~E. Harris, \emph{Theory of Branching Processes}, Dover NY, 1989.

\bibitem{em}
A. P. Dempster, N. M. Laird, and D. B. Rubin, ``Maximum likelihood from incomplete data via the EM algorithm,"
\emph{Journal of the Royal Statistical Society. Series B (Methodological)}, vol. 39, no. 1, pp. 1--38, Jan. 1977.

\bibitem{good}
I. J. Good, ``Generalizations to several variables of Lagrange's expansion, with applications to stochastic processes," \emph{Mathematical Proceedings of the Cambridge Philosophical Society}, vol. 56, pp. 367--380, 1960.

\bibitem{inter}
S. V. Buldyrev, R. Parshani, G. Paul, H. E. Stanley, and S. Havlin, ``Catastrophic cascade of failures in interdependent network,'' \emph{Nature}, vol. 464, no. 7291,
pp. 1025--1028, Apr. 2010.

\bibitem{commu}
M. Parandehgheibi, E. Modiano, and D. Hay, ``Mitigating cascading failures in interdependent power grids and communication networks,'' 
\emph{IEEE Intl. Conf. Smart Grid Comm. (SmartGridComm)}, pp. 242--247, Nov. 2014.

\bibitem{gas1}
M. Shahidehpour, Y. Fu, and T. Wiedman, ``Impact of natural gas infrastructure on electric power systems,'' \emph{Proc. IEEE}, vol. 93, no. 5, pp. 1042--1056, May 2005.

\bibitem{gas}
T. Li, M. Eremia, and M. Shahidehpour, ``Interdependency of natural gas network and power system security,'' 
\emph{IEEE Trans. Power Syst.}, vol. 23. no. 4, pp. 1817--1824, Nov. 2008.

\bibitem{water}
T. Adachi and B. R. Ellingwood, ``Serviceability of earthquake-damaged water systems: Effects of electrical power availability and power backup systems 
on system vulnerability,'' \emph{Reliability Engineering and System Safety}, vol. 93, no. 1, pp. 78--88, Jan. 2008.

\bibitem{transport}
M. Amin, ``Toward self-healing infrastructure systems,'' \emph{Computer}, vol. 8, pp. 44--53, 2000.

\bibitem{transport1}
S. M. Rinaldi, J. P. Peerenboom, T. K. Kelly, ``Identifying, understanding, and analyzing critical infrastructure interdependencies,'' \emph{IEEE Control Systems}, vol. 21, no. 6, pp. 11--25, Dec. 2001.

\bibitem{Guttorp}
P. Guttorp, \emph{Statistical Inference for Branching Processes}, Wiley NY, 1991.

\bibitem{estimator}
F. Maaouia and A. Touati, ``Identification of multitype branching processes," \emph{The Annals of Statistics}, vol. 33, no. 6, pp. 2655--2694, 2005.


\end{thebibliography}
\end{document}